\documentclass[journal,twocolumn,10pt]{IEEEtran}

\usepackage{mathrsfs}

\usepackage{amssymb}

\usepackage[mathcal]{euscript}

\usepackage{psfrag,calc,url,bm}

\usepackage{cite}

\usepackage{graphicx}

\usepackage{psfrag}

\usepackage{subfigure}

\usepackage{url}

\usepackage{stfloats}

\usepackage{amsmath}

\usepackage{float}

\usepackage{algorithm}

\usepackage{algorithmic}

\usepackage{color}
\begin{document}

\title{Group Cooperation with Optimal Resource Allocation in Wireless Powered Communication Networks}

\author{Ke~Xiong,~\IEEEmembership{Member,~IEEE,} Chen Chen, Gang Qu,~\IEEEmembership{Senior Member, ~IEEE}\\ Pingyi Fan,~\IEEEmembership{Senior Member,~IEEE},~~Khaled Ben Letaief,~\IEEEmembership{Fellow,~IEEE}\\

\thanks{%Manuscript received xxx, 2015; revised xxx,  and
%accepted xxx. Date of publication June xxx;
%date of current version xxx. This work was supported by the National
%Basic Research Program of China (973 Program), no. 2012CB316100(2) and also by the Conhealth Project, no. 294923.
Ke~Xiong is with the School of Computer and Information Technology,  Beijing Jiaotong University,  Beijing 100044,  R.P. China. e-mail:  kxiong@bjtu.edu.cn.

Chen Chen and Gang Qu are with Department of Electrical \& Computer Engineering, University of Maryland, College Park, USA. email: ccmmbupt@gmail.com, Gangqu@umd.edu.

Pingyi Fan is with the Department of Electronic Engineering,  Tsinghua University,  Beijing,  R.P. China, 100084. e-mail:  fpy@tsinghua.edu.cn.

K. B. Letaief is with the School of Engineering, Hong Kong University of Science \& Technology (HKUST), China. e-mail:  eekhaled@ece.ust.hk.
}
}
%\author{Chen Chen$^{\star}$, Ke Xiong$^{\dagger,\natural}$, Gang Qu$^{\star}$, Pingyi Fan$^\sharp$\\
%\small
%$^{\star}$Department Of Electrical \& Computer Engineering, University Of Maryland, College Park, USA\\
%$^\dagger$School of Computer and Information Technology, Beijing Jiaotong University, Beijing, China\\
%$^\natural$National Mobile Communications Research Laboratory, Southeast University, Nanjing, China\\
%$^\sharp$Department of Electronics Engineering,  Tsinghua University,  Beijing, China\\
%kxiong@bjtu.edu.cn}

\maketitle

\begin{abstract}
This paper considers a wireless powered communication network (WPCN) with group cooperation, where two communication groups cooperate with each other via wireless power transfer and time sharing to fulfill their expected information delivering \textcolor[rgb]{0.00,0.00,0.00}{and achieve ``win-win'' collaboration}. To explore the system performance limits, we formulate optimization problems to respectively maximize the weighted sum-rate (WSR) and minimize the total consumed power. The time assignment, beamforming vector and power allocation are jointly optimized under available power and quality of service (QoS) requirement constraints of both groups. For the WSR-maximization, both fixed and flexible power scenarios are investigated. As all problems are non-convex and have no known solution methods, we solve them by using proper variable substitutions and the semi-definite relaxation (SDR). We theoretically prove that our proposed solution method guarantees the global optimum for each problem. Numerical results are presented to show the system performance behaviors, which provide some useful insights for future WPCN design. \textcolor[rgb]{0.00,0.00,0.00}{It shows that in such a group cooperation-aware WPCN, optimal time assignment has the most great effect on the system performance than other factors.}
\end{abstract}


\section{Introduction}
%\subsection{Background}
Recently, the fast development of radio frequency (RF)-based wireless power transfer (WPT) technology\cite{RF2,avullers,Sbiho}  makes it possible to build wireless
powered communication networks (WPCNs)\cite{Sbiho,Suzhibi}, in which  \textcolor[rgb]{0.00,0.00,0.00}{communication devices} can be remotely powered over the air by
dedicated wireless energy transmitters.  Compared with traditional battery-powered
networks, WPCN avoids the manual battery replacement/recharging, which reduces the network maintenance and operation cost greatly. As the transmit power, waveforms, and occupied time/frequency
dimensions, etc., of WPT are all controllable and tunable, it is capable of providing stable energy supply under various physical conditions and communication requirements in WPCNs\cite{akjrliu,aulukus,akrikidiss}.

It was reported that tens of micowatts RF power can be transferred to a distance of more than 10 meters by using RF-based WPT\cite{Suzhibi}.  The energy is sufficient to power the low-power communication devices (e.g., sensors and RF identification (RFID) tags). Thanks to the rapid evolution of multi-antenna energy beamforming\cite{Slee}, high-efficiency energy harvesting (EH) circuit design\cite{Georgiadis} and energy efficient communication system design\cite{Keee}, RF-based WPT has been regarded \textcolor[rgb]{0.00,0.00,0.00}{as a} promising and attractive solution to prolong the lifetime of low-power energy-constrained networks, such as wireless sensor networks (WSNs), wireless body area networks (WBANs) and Internet of Things (IoT) in future 5G systems\cite{Chen,zZHOU,Qzyao,SZbi,Suzhibi,Slee,Keee}.

Since RF signals also carry energy when they transfer information, simultaneous
wireless information and power transfer (SWIPT) technology  was proposed\cite{Varshney,Grover,Rzhang}, which has captured greatly attention.
%The pioneering works on SWIPT were \cite{Varshney} and \cite{Grover}, which investigated the fundamental tradeoff between the capacity and harvested energy for SWIPT systems. Later, several practical receiver architectures were proposed in \cite{Rzhang}, where the the rate-energy region was derived for multi-input multi-output (MIMO) broadcasting SWIPT systems.
It was proved that SWIPT is more
efficient in spectrum usage than transmitting information and energy in orthogonal time/
frequency/spacial channels\cite{Rzhang,ke,XZhou,SWIPT4,SWIPT5,SWIPT6,SWIPT7}.

%\subsection{Related Work}
So far, SWIPT-enabled WPCNs have been attracting increasing interests, see e.g. \textcolor[rgb]{0.00,0.00,0.00}{\cite{JuThroughput,JuUser,YLche,QQwu,Fzhao,YLchexu,Liu,Sun,Yyma}.  In \cite{JuThroughput,JuUser,YLche,QQwu,Fzhao}, single-antenna hybrid access point (H-AP)-assisted WPCN was investigated, where the system throughput or weighted sum-rate (WSR) were maximized via optimal time assignments. Since only single antenna was assumed at the H-AP, no beamforming design was involved in their works. As is known, with multiple antennas equipped at the transmitter, beamforming can be employed improve the energy/infromation transmission efficiency due to its focusing effect of the signals on specific receivers. Thus, some works began to consider beamforming design in WPCNs, see e.g., \cite{YLchexu,Liu,Sun}.
In \cite{YLchexu}, beamforming vectors were optimized to maximizing the system achievable information rate.
In \cite{Liu} and \cite{Sun}, beamforming vectors were jointly optimized with time assignment to maximize the sum-rate of the WPCN with a multi-antenna H-AP. Seeing that WPCN provides a promising solution for WSN and IoT, in which information is often relayed over multiple hops from a source to its destination due to the limited coverage of each node, some works also investigated WPCN with relay technologies, see e.g. \cite{ke} and \cite{Yyma}, where amplify-and-forward (AF) and decode-and-forward (DF) relay operations were studied in \cite{ke} and \cite{Yyma}, respectively. Besides, some existing works also investigated the resource allocation of WPCN in various wireless networks, see e.g. \cite{Hju, Hjkim,HoonLee}.}

%\subsection{Motivation \& Contributions}
\textcolor[rgb]{0.00,0.00,0.00}{However, existing works only studied the energy transfer and information delivering within the same communication group, which means that the energy was transferred from the H-AP to its users and the users used the harvested energy to transmit information to the H-AP or  the energy was transferred from the source to the energy constrained relay node and then the relay help to forward the information from the source to its destinations. Therefore, no group cooperation was involved in exsting works and the systems were designed only by considering the utility maximization of the single communication group.}

In this paper, we investigate the group cooperation with optimal resource allocation in WPCNs. We consider a network composed of two communication groups, where the group 1 has sufficient \textcolor[rgb]{0.00,0.00,0.00}{energy} supply but no licensed bandwidth, and the  group 2 has licensed bandwidth but no sufficient \textcolor[rgb]{0.00,0.00,0.00}{energy}. Therefore, neither group can fulfill the information delivering  to meet its desired information transmission rate.
Considering that SWIPT provides an effective approach for information transmission and energy cooperation between nodes, we introduce the \textcolor[rgb]{0.00,0.00,0.00}{energy cooperation and time sharing} between the two groups, so that group 1 may transfer some \textcolor[rgb]{0.00,0.00,0.00}{energy} to group 2 and then get some \textcolor[rgb]{0.00,0.00,0.00}{transmission time} from group 2 in return. With this inter-group cooperation, both groups can achieve their expected information rates.
For such a WPCN with group cooperation, our goal is to explore its performance limits in terms of WSR and the minimum consumed power.

\textcolor[rgb]{0.00,0.00,0.00}{Compared with existing works, \textbf{several other differences}} of our work are emphasized as follows. \textcolor[rgb]{0.00,0.00,0.00}{\textit{\textbf{Firstly}}}, different from some existing works on one-hop WPCNs, see e.g., \cite{JuThroughput,YLchexu},  where only point-to-point communication was investigated, in our work, cooperative relaying \footnote{\textcolor[rgb]{0.00,0.00,0.00}{In our work, DF relaying cooperation is employed since DF relaying often outperforms AF relaying, especially in relatively high signal-to-noise ratio (SNR) scenarios.}} is involved. Although some works studied the  relay-aided WPCN systems, see e.g. \cite{zZHOU,Yyma}, \textcolor[rgb]{0.00,0.00,0.00}{all nodes were assumed with single antenna, so that no beamforming was considered in their work.}
\textcolor[rgb]{0.00,0.00,0.00}{\emph{\textbf{Secondly}}, although some works introduced cooperation into WPCNs, they did not investigate the ``win-win'' collaboration via energy and time cooperation between different groups. For example, in \cite{JuUser}, the user cooperation was studied in relay-aided WPCN, where the closer user was powered to help the farther user forward information. However, no energy transfer cooperation between the two users was involved and no beamforming was considered. In \cite{Sanket}, the cooperation between the primary users and secondary users in cognitive networks was studied, where however, only the sum-rate of the secondary users was maximized and the beamforming design also was not involved. Comparably, in our work, the group cooperation in terms of wireless power transfer and time sharing are involved to achieve a ``\emph{win-win}'' collaboration and the SWIPT beamforming is also considered.} \textcolor[rgb]{0.00,0.00,0.00}{\textit{\textbf{Thirdly}}, different from most existing works, see e.g, \cite{QQwu,Yyma,Qzyao,SZbi}, where only one or two kinds of resources were optimized, in our work, cooperative relaying, time assignment, SWIPT beamforming and power allocation with group cooperation are jointly designed and optimized in a single system and we mathematically prove that our proposed optimization method achieves the global optimum.}

The \emph{contributions} of our work are summarized as follows.

\emph{Firstly}, we propose a group-cooperation based cooperative transmission protocol for the considered WPCN, \textcolor[rgb]{0.00,0.00,0.00}{which is able to achieve ``win-win" cooperation transmission between two communication groups via energy transfer and time sharing.}

\emph{Secondly}, to explore the information transmission performance limit of the system,  we formulate two optimization problems to maximize the system WSR by jointly optimizing the time assignment and beamforming vector under two different power constraints, i.e., the fixed power and the flexible power constraints. In order to achieve the ``win-win" cooperation between the two groups and guarantee their QoS requirements, the minimal required information rate constraints of the two groups are also considered in the optimal system design. As both problems are non-convex and have no known solution methods, we transform them into equivalently ones with some variable substitutions and then solve them by using semi-definite relaxation (SDR) method.  We theoretically prove that our proposed solution method can guarantee to find the global optimal solution.

\emph{Thirdly}, consider that WPCNs have promising application potentials in future energy-constrained networks, in which the power consumption reduction is very critical and the green communication design \cite{akjrliu,aulukus,akrikidiss,Keee,QQwu,EE} is very essential. We formulate an optimization problem to minimize the total consumed power of the WPCN by jointly optimizing the time assignment and beamforming vector under required data rate constraints of the two groups. As the problem is non-convex, we also solve it efficiently by using some variable substitutions and the SDR method.  The global optimum of our proposed minimal power consumption system design is also theoretically proved.

\textit{Fourthly}, numerical results are presented to discuss the system performance behaviors, which provide some useful insights for future WPCN design. It shows that the average power constrained system achieves higher WSR than the fixed power constrained system \textcolor[rgb]{0.00,0.00,0.00}{and in such a group cooperation-aware WPCN, optimal time assignment has the most great effect on the system performance than other factors. Besides, the effects of relay position on system performances are also discussed via simulations.}

%\subsection{Organization}
The rest of the paper is organized as follows. Section II describes the system model. Section \textcolor[rgb]{0.00,0.00,0.00}{III} and IV investigate the  WSR maximization and power minimization design of our considered WPCN, respectively. Section V provides some simulation results and finally, Section VI concludes the paper.

\section{System Model}\label{Sec:symodel}

\subsection{Network Model}
\begin{figure}
\centering
\includegraphics[width=0.475\textwidth]{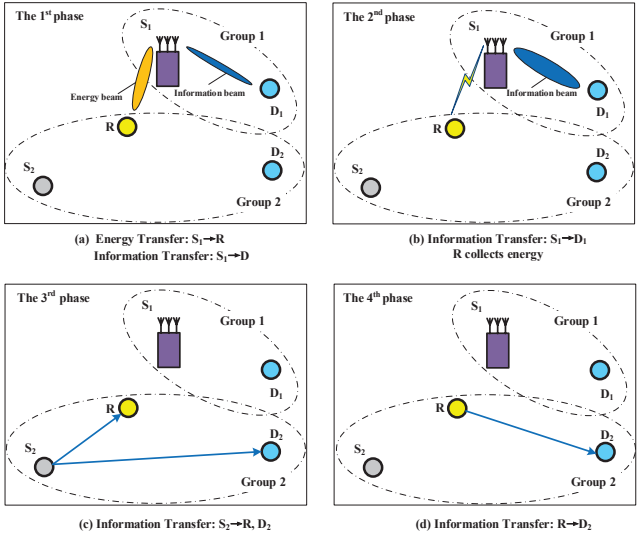}
\caption{\textcolor[rgb]{0.00,0.00,0.00}{System model and the 4-phase cooperative transmission protocol.}}
\label{fig:sysm}
\end{figure}

Consider a wireless system consisting of two communication groups as shown in Figure \ref{fig:sysm}, where  in group 1 source node $\rm{S}_1$ desires to transmit information to $\rm{D}_1$ and in group 2 source node $\rm{S}_2$ desires to transmit information to $\rm{D}_2$. For group 1, $\rm{S}_1$ is with stable and sufficient \textcolor[rgb]{0.00,0.00,0.00}{energy} supply but no licensed bandwidth, so it cannot transmit information to $\rm{D}_1$. For group 2, $\rm{S}_2$ has licensed bandwidth but it is located relatively far away from $\rm{D}_2$, so it cannot achieve high enough data rate over $\rm{S}_2 \rightarrow \rm{D}_2$ direct link to meet its required information rate. Thus, $\rm{S}_2$ needs $\rm{R}$ to help it forward information to $\rm{D}_2$. It is assumed that $\rm{R}$ is an energy-exhausted/selfish node, so that $\rm{R}$ cannot or is not willing to consume its own \textcolor[rgb]{0.00,0.00,0.00}{energy} to help the information forwarding from $\rm{S}_2$ to $\rm{D}_2$. In this case, neither group 1 (i.e., the bandwidth-limited group) nor group 2 (i.e., the power-limited group) can fulfill its expected information delivery.

\textcolor[rgb]{0.00,0.00,0.00}{Fortunately}, by using WPT, the two groups \textcolor[rgb]{0.00,0.00,0.00}{is able to} cooperate with each other in terms of energy and \textcolor[rgb]{0.00,0.00,0.00}{transmission time} to achieve a ``win-win'' outcome to fulfill their respectively desired information transmission. Specifically, $\rm{S}_1$  transmits some \textcolor[rgb]{0.00,0.00,0.00}{energy} to $\rm{R}$ to enable $\rm{R}$ participating in the information transmission from $\rm{S}_2$ to $\rm{D}_2$. In return, $\rm{S}_2$ bestows a portion of its transmission time to $\rm{S}_1$  to help group 1 accomplish the information delivery. With such a cooperation, both groups, therefore, may successfully deliver their information.

\textcolor[rgb]{0.00,0.00,0.00}{It is marked that our presented cooperation model  can also be applied in cognitive radio networks, where group 2 can be regarded as the primary user with listened frequency band and group 1 can be regarded as secondary users with no licensed frequency band. In traditional underlay cognitive networks, group 1 transmits information only when group 2 is silent. If group 2 always transmits signals, group 1 has no opportunity to transmit its information. Besides, due to the week direct link in group 1, its achievable information rate may be pretty low. However, with our described energy and time sharing cooperation, group 1 is motivated to share its transmission time with group 2 and is able to get some energy to increase its information rate. Meanwhile, group 2 will not to passively wait for a chance to transmit its information and it can actively seek some transmission opportunity at the expense of some energy. Therefore, the underlay cognitive transmission of the primary and the secondary users, as two cooperative groups, could obtain their profits.}

In order to enhance the energy transfer efficiency, $\rm{S}_1$ (\textcolor[rgb]{0.00,0.00,0.00}{e.g. a sink node in WSN}) is assumed to be equipped with $N$ antennas while all other nodes (\textcolor[rgb]{0.00,0.00,0.00}{e.g. sensor nodes}) only support single antenna due to their size limitations. Block fading channel is considered, so that all channel coefficients can be regarded as constants during each fading block and vary from block to block independently, following Rayleigh distribution. $h_{u v}(k)$ is used to denote the channel coefficient of the $k$-th block between node $u$ and node $v$. $n(k) \sim \mathcal{CN} (0, N_0)$ is the Additive White Gaussian Noise (AWGN) of the $k$-th block. So, $h_{u v}(k) \sim \mathcal{CN} (0, d_{u v}^{-\beta})$, where $d_{u v}$ is the distance between node $u$ and node $v$, and $\beta$ is the path loss exponent factor. The time period of each fading block is denoted by $T$.

\subsection{Transmission Protocol}
To complete cooperation transmission, each time period $T$ is divided into four phases, which are with time intervals of $\tau_1$, $\textcolor[rgb]{0.00,0.00,0.00}{\tau_2}$, $\textcolor[rgb]{0.00,0.00,0.00}{\tau_3}$ and $\textcolor[rgb]{0.00,0.00,0.00}{\tau_4}$, respectively, where $\tau_m \geq 0$ with $m=1,2,...,4$. Without loss of generality, $T$ is normalized to 1 in the sequel, so that $\sum\nolimits_{m=1}^4 \tau_m = 1$. Defining $\bm{\tau} \triangleq [\tau_1\,\,\textcolor[rgb]{0.00,0.00,0.00}{\tau_2}\,\,\textcolor[rgb]{0.00,0.00,0.00}{\tau_3}\,\,\textcolor[rgb]{0.00,0.00,0.00}{\tau_4}]^T$ as the time assignment vector of the four transmission phases, it satisfies that
\begin{equation}\label{eqn:tau}
\mathbf{1}^T \bm{\tau} = 1,\,\,\bm{\tau} \succeq \mathbf{0},
\end{equation}
where $\mathbf{1}$ is a column vector with all elements being 1.

In the first phase with time interval $\tau_1$, $\rm{S}_1$ transfers energy to $\rm{R}$ and transmits information to $\rm{D}_1$ simultaneously. Let $x_{\rm{S}_1}(k)$ with $|x_{\rm{S}_1}(k)|^2 = 1$ be the transmitted symbol from $\rm{S}_1$. The received signals at $\rm{D}_1$ and $\rm{R}$ are, respectively, given by
\begin{equation}
y_{\rm{D}_1}(k) = \sqrt{P_{\rm{S}_1}^{(1)}}\mathbf{h}_{\rm{S}_1 \rm{D}_1}^H (k)\bm{\omega} x_{\rm{S}_1}(k) + n(k)
\end{equation}
and
\begin{equation}
y_{\rm R}(k) = \sqrt{P_{\rm{S}_1}^{(1)}}\mathbf{h}_{\rm{S}_1 \rm{R}}^H (k)\bm{\omega} x_{\rm{S}_1}(k) + n(k) ,
\end{equation}
where $\mathbf{h}_{\rm{S}_1 \rm{D}_1}\in \mathbb{C}^{N \times 1}$ and $\mathbf{h}_{\rm{S}_1 \rm{R}}\in \mathbb{C}^{N \times 1}$ are the complex channel vectors from $\rm{S}_1$ to  $\rm{D}_1$ and from $\rm{S}_1$ to $\rm{R}$, respectively.  $P_{\rm{S}_1}^{(1)}$ is the available transmit power at $\rm{S}_1$ in the first phase. $\bm{\omega} \in \mathbb{C}^{N \times 1}$ represents the beamforming vector at $\rm{S}_1$, satisfying
\begin{equation}\label{eqn:wnorm}
\| \bm{\omega}\|^2 \leq 1.
\end{equation}
The achievable information rate in the first phase at $\rm{D}_1$ can be given by
\begin{equation}
R_{\rm{S}_1}^{(1)} = \tau_1\mathcal{C}\left(\frac{P_{\rm{S}_1}^{(1)} |\mathbf{h}_{\rm{S}_1 \rm{D}_1}^H \bm{\omega}|^2}{N_0}\right),
\end{equation}
where \textcolor[rgb]{0.00,0.00,0.00}{$\mathcal{C}(x) \triangleq \log_2(1 + x)$} and the harvested energy at $\rm{R}$ is
\begin{equation}\label{eqn:er}
E_{\rm R}^{\textcolor[rgb]{0.00,0.00,0.00}{(1)}} = \eta \tau_1 P_{\rm{S}_1}^{(1)}| \mathbf{h}_{\rm{S}_1 \rm{R}}^H \bm{\omega} |^2,
\end{equation}
where $\eta\in(0, 1]$ is a constant, accounting for the energy conversion efficiency. The larger the value of $\eta$, the higher the energy conversion efficiency. In particular, $\eta = 1$ means all received signal power can be perfectly converted to energy at the receiver.

\textcolor[rgb]{0.00,0.00,0.00}{In the second phase, with time interval $\textcolor[rgb]{0.00,0.00,0.00}{\tau_2}$ rewarded by group 2, $\rm{S}_1$ transmits its own information to $\rm{D}_1$ via multiple antennas.} As it is a typical multiple input single output (MISO) channel, by using \textcolor[rgb]{0.00,0.00,0.00}{the maximum rate transmission (MRT) strategy\cite{DTse}}, the achievable information rate from $\rm{S}_1$ to $\rm{D}_1$ in this phase can by given by
\begin{equation}
R_{\rm{S}_1}^{\textcolor[rgb]{0.00,0.00,0.00}{(2)}} = \textcolor[rgb]{0.00,0.00,0.00}{\tau_2}\mathcal{C}\left(\frac{P_{\rm{S}_1}^{\textcolor[rgb]{0.00,0.00,0.00}{(2)}} \|\mathbf{h}_{\rm{S}_1 \rm{D}_1}\|^2}{N_0}\right),
\end{equation}
where $P_{\rm{S}_1}^{\textcolor[rgb]{0.00,0.00,0.00}{(2)}}$ is the available transmit power at $\rm{S}_1$ in the \textcolor[rgb]{0.00,0.00,0.00}{second} phase. \textcolor[rgb]{0.00,0.00,0.00}{Because of the broadcast nature of wireless channel, in this phase, the transmitted signals from $\rm{S}_1$ also can be collected by $\rm{R}$ for energy harvesting. So, the harvested energy in the second phase can be given by}
\textcolor[rgb]{0.00,0.00,0.00}{
\begin{equation}\label{eqn:er2}
E_{\rm R}^{(2)}= \eta \tau_2 P_{\rm{S}_1}^{(2)}| \mathbf{h}_{\rm{S}_1 \rm{R}}^H \tfrac{\mathbf{h}_{\rm{S}_1 \rm{D}_1}}{\parallel\mathbf{h}_{\rm{S}_1 \rm{D}_1}\parallel}|^2,
\end{equation}
where $\tfrac{\mathbf{h}_{\rm{S}_1 \rm{D}_1}}{\parallel\mathbf{h}_{\rm{S}_1 \rm{D}_1}\parallel}$ is the transmission precoding vector adopted at $\rm{S}_1$ for MRT.}

In the \textcolor[rgb]{0.00,0.00,0.00}{third} phase with time interval $\textcolor[rgb]{0.00,0.00,0.00}{\tau_3}$, $\rm{S}_2$ broadcasts information to $\rm{R}$ and $\rm{D}_2$. Let the transmitted symbol by $\rm{S}_2$ be $x_{\rm{S}_2}(k)$ with $|x_{\rm{S}_2}(k)|^2 = 1$. The signal received at $\rm{R}$ and $\rm{D}_2$ can be, respectively, given by
\begin{equation}
y_{\rm R}(k) = \sqrt{P_{\rm{S}_2}^{\textcolor[rgb]{0.00,0.00,0.00}{(3)}}} h_{\rm{S}_2 \rm{R}}(k) x_{\rm{S}_2}(k) + n(k)
\end{equation}
and
\begin{equation}
y_{\rm{D}_2}(k) = \sqrt{P_{\rm{S}_2}^{\textcolor[rgb]{0.00,0.00,0.00}{(3)}}} h_{\rm{S}_2 \rm{D}_2}(k) x_{\rm{S}_2}(k) + n(k),
\end{equation}
where $P_{\rm{S}_2}^{\textcolor[rgb]{0.00,0.00,0.00}{(3)}}$ is the available transmit power at $\rm{S}_2$.

In the \textcolor[rgb]{0.00,0.00,0.00}{fourth} phase with time interval $\textcolor[rgb]{0.00,0.00,0.00}{\tau_4}$, $\rm{R}$ decodes the information transmitted from $\rm{S}_2$ and then helps to forward the decoded information to $\rm{D}_2$ by using the harvested energy from $\rm{S}_1$ in the \textcolor[rgb]{0.00,0.00,0.00}{first two phases}. The received signal at $\rm{D}_2$ from $\rm{R}$ in the third phase is
\begin{equation}
y_{\rm{D}_2}(k) = \sqrt{P_{\rm{R}}} h_{\rm{R} \rm{D}_2}(k) x_{R}(k) + n(k) ,
\end{equation}
where $P_{\rm{R}}$ is the available transmit power at $\rm{R}$, which is constrained by \textcolor[rgb]{0.00,0.00,0.00}{the sum of the harvested energy in the first two phases, i.e., $E_{\rm R}^{(1)}$ in (\ref{eqn:er}) and $E_{\rm R}^{(2)}$ in (\ref{eqn:er2})}. That is,
\begin{flalign}\label{eqn:renergy}
\textcolor[rgb]{0.00,0.00,0.00}{\tau_4} P_{\rm{R}} &\leq E_{\rm R}^{(1)}+E_{\rm R}^{(2)}\\
&= \eta \tau_1 P_{\rm{S}_1}^{(1)}| \mathbf{h}_{\rm{S}_1 \rm{R}}^H \bm{\omega} |^2 \textcolor[rgb]{0.00,0.00,0.00}{+\eta \tau_2 P_{\rm{S}_1}^{(2)}| \mathbf{h}_{\rm{S}_1 \rm{R}}^H \tfrac{\mathbf{h}_{\rm{S}_1 \rm{D}_1}}{\parallel\mathbf{h}_{\rm{S}_1 \rm{D}_1}\parallel}|^2.}\nonumber
\end{flalign}
Decode-and-forward (DF) relaying operation is employed at $\rm{R}$, so the end-to-end information rate of group 2 satisfies that \cite{DF}
%\begin{flalign}\label{eqn:rs2}
%R_{\rm{S}_2} = \min \Bigg\{&\textcolor[rgb]{0.00,0.00,0.00}{\tau_3}\mathcal{C}\left(\frac{P_{\rm{S}_2}^{\textcolor[rgb]{0.00,0.00,0.00}{(3)}} |h_{\rm{S}_2 \rm{R}}|^2}{N_0}\right),\\ &\textcolor[rgb]{0.00,0.00,0.00}{\tau_3}\mathcal{C}\left(\frac{P_{\rm{S}_2}^{\textcolor[rgb]{0.00,0.00,0.00}{(3)}} |h_{\rm{S}_2 \rm{D}_2}|^2}{N_0}\right)+\textcolor[rgb]{0.00,0.00,0.00}{\tau_4}\mathcal{C}\left(\frac{P_{\rm{R}} |h_{\rm{R} \rm{D}_2}|^2}{N_0}\right) \Bigg\}.\nonumber
%\end{flalign}
\begin{flalign}\label{eqn:rs2}
R_{\rm{S}_2} \textcolor[rgb]{0.00,0.00,0.00}{\leq} \min \Bigg\{&\textcolor[rgb]{0.00,0.00,0.00}{\tau_3}\mathcal{C}\left(\frac{P_{\rm{S}_2}^{\textcolor[rgb]{0.00,0.00,0.00}{(3)}} |h_{\rm{S}_2 \rm{R}}|^2}{N_0}\right),\\ &\textcolor[rgb]{0.00,0.00,0.00}{\tau_3}\mathcal{C}\left(\frac{P_{\rm{S}_2}^{\textcolor[rgb]{0.00,0.00,0.00}{(3)}} |h_{\rm{S}_2 \rm{D}_2}|^2}{N_0}\right)+\textcolor[rgb]{0.00,0.00,0.00}{\tau_4}\mathcal{C}\left(\frac{P_{\rm{R}} |h_{\rm{R} \rm{D}_2}|^2}{N_0}\right) \Bigg\}.\nonumber
\end{flalign}

For the four phases described above, group 1 transmits information in both the first and the \textcolor[rgb]{0.00,0.00,0.00}{second} phases. Combining $R_{\rm{S}_1}^{(1)}$ with $R_{\rm{S}_1}^{\textcolor[rgb]{0.00,0.00,0.00}{(2)}}$, one can obtain the total achievable information rate from $\rm{S}_1$ to $\rm{D}_1$ in the $k$-th fading block as
%\begin{flalign}\label{eqn:rs1}
%R_{\rm{S}_1} &= R_{\rm{S}_1}^{(1)} + R_{\rm{S}_1}^{\textcolor[rgb]{0.00,0.00,0.00}{(2)}}\\
%&= \tau_1\mathcal{C}\left(\frac{P_{\rm{S}_1}^{(1)}|\mathbf{h}_{\rm{S}_1 \rm{D}_1}^H \bm{\omega}|^2}{N_0}\right)
%+ \textcolor[rgb]{0.00,0.00,0.00}{\tau_2}\mathcal{C}\left(\frac{P_{\rm{S}_1}^{\textcolor[rgb]{0.00,0.00,0.00}{(2)}} \|\mathbf{h}_{\rm{S}_1 \rm{D}_1}\|^2}{N_0}\right).\nonumber
%\end{flalign}
\begin{flalign}\label{eqn:rs1}
R_{\rm{S}_1}&\leq R_{\rm{S}_1}^{(1)} + R_{\rm{S}_1}^{\textcolor[rgb]{0.00,0.00,0.00}{(2)}}
\textcolor[rgb]{0.00,0.00,0.00}{=}\\
&\tau_1\mathcal{C}\left(\frac{P_{\rm{S}_1}^{(1)}|\mathbf{h}_{\rm{S}_1 \rm{D}_1}^H \bm{\omega}|^2}{N_0}\right)
+ \textcolor[rgb]{0.00,0.00,0.00}{\tau_2}\mathcal{C}\left(\frac{P_{\rm{S}_1}^{\textcolor[rgb]{0.00,0.00,0.00}{(2)}} \|\mathbf{h}_{\rm{S}_1 \rm{D}_1}\|^2}{N_0}\right).\nonumber
\end{flalign}
Group 2 transmits information in the \textcolor[rgb]{0.00,0.00,0.00}{third and the fourth} phases via the DF cooperative relaying, whose available information rate in the $k$-th fading block is given by (\ref{eqn:rs2}).

Suppose the minimal required information rate of group $i$ is $r_{\rm{S}_i}$, where $i \in \{1, 2\}$. The end-to-end achievable information rate $R_{\rm{S}_i}$ satisfies that
\begin{equation}\label{eqn:r12}
  R_{\rm{S}_i} \geq r_{\rm{S}_i},\,\, \forall i = 1, 2.
\end{equation}
Note that the minimal required data rate constraints in (\ref{eqn:r12}) are reasonable and practical in the considered WPCN system, because only when the obtained data rates exceed the minimal required ones, the cooperation between the two groups brings benefits to both groups. \textcolor[rgb]{0.00,0.00,0.00}{Also, with the minimal required data rate constraints in (\ref{eqn:r12}), the problems in section III may not have feasible solution. In this case, it indicates that there is no opportunity for the two groups to achieve win-win cooperation.}

\section{WSR-maximization Design}
Let $\alpha_i \geq 0$ be the weight of achievable information rate of group $i$, where $i=1,2$. The WSR of the system can be given by
\begin{equation}
  R_{\rm{wsum}} = \alpha_1 R_{\rm{S}_1} + \alpha_2 R_{\rm{S}_2}.
\end{equation}
\textcolor[rgb]{0.00,0.00,0.00}{We shall consider two different scenarios, i.e., the fixed and the flexible power scenarios, for the WSR-maximization design of the cooperative WPCN in the following two subsections.}

\subsection{Fixed Power Scenario}
\subsubsection{Problem Formulation}
In the fixed power scenario, ${\rm S}_1$ and ${\rm S}_2$ have fixed instantaneous powers in their respective transmission phases. \textcolor[rgb]{0.00,0.00,0.00}{For ${\rm S}_1$ it uses the same transmit power to transmit signals \textcolor[rgb]{0.00,0.00,0.00}{in phase 1 and phase 2}, i.e., $P_{{\rm S}_1}^{(1)} = P_{{\rm S}_1}^{\textcolor[rgb]{0.00,0.00,0.00}{(2)}}$. For ${\rm S}_2$ it transmits signals \textcolor[rgb]{0.00,0.00,0.00}{in phase 3} with the transmit power $P_{{\rm S}_2}^{\textcolor[rgb]{0.00,0.00,0.00}{(3)}}$.} For clarity, we denote the fixed power at ${\rm S}_i$ to be $P_{{\rm S}_i}$, so we have that $P_{{\rm S}_1}^{(1)} = P_{{\rm S}_1}^{\textcolor[rgb]{0.00,0.00,0.00}{(2)}} = P_{{\rm S}_1}$ and $P_{{\rm S}_2}^{\textcolor[rgb]{0.00,0.00,0.00}{(3)}} = P_{{\rm S}_2}$. As a result, (\ref{eqn:rs2}) and (\ref{eqn:rs1}) can be respectively rewritten as
%\begin{flalign}\label{eqn:rs2fp}
%R_{{\rm S}_2} = \min \Bigg\{&\textcolor[rgb]{0.00,0.00,0.00}{\tau_3}\mathcal{C}\left(\frac{P_{{\rm S}_2} |h_{{\rm S}_2 \rm{R}}|^2}{N_0}\right),\\ &\textcolor[rgb]{0.00,0.00,0.00}{\tau_3}\mathcal{C}\left(\frac{P_{{\rm S}_2} |h_{{\rm S}_2 {\rm D}_2}|^2}{N_0}\right)+\textcolor[rgb]{0.00,0.00,0.00}{\tau_4}\mathcal{C}\left(\frac{P_{\rm R} |h_{{\rm R} {\rm D}_2}|^2}{N_0}\right) \Bigg\},\nonumber
%\end{flalign}
\begin{flalign}\label{eqn:rs2fp}
R_{{\rm S}_2} \textcolor[rgb]{0.00,0.00,0.00}{\leq} \min \Bigg\{&\textcolor[rgb]{0.00,0.00,0.00}{\tau_3}\mathcal{C}\left(\frac{P_{{\rm S}_2} |h_{{\rm S}_2 \rm{R}}|^2}{N_0}\right),\\
&\textcolor[rgb]{0.00,0.00,0.00}{\tau_3}\mathcal{C}\left(\frac{P_{{\rm S}_2} |h_{{\rm S}_2 {\rm D}_2}|^2}{N_0}\right)+\textcolor[rgb]{0.00,0.00,0.00}{\tau_4}\mathcal{C}\left(\frac{P_{\rm R} |h_{{\rm R} {\rm D}_2}|^2}{N_0}\right) \Bigg\},\nonumber
\end{flalign}
and
\begin{flalign}\label{eqn:rs1fp}
R_{{\rm S}_1} \textcolor[rgb]{0.00,0.00,0.00}{\leq} \tau_1\mathcal{C}\left(\frac{P_{\rm{S}_1}|\mathbf{h}_{{\rm S}_1 {\rm D}_1}^H \bm{\omega}|^2}{N_0}\right)
+ \textcolor[rgb]{0.00,0.00,0.00}{\tau_2}\mathcal{C}\left(\frac{\textcolor[rgb]{0.00,0.00,0.00}{P_{{\rm S}_1}} \|\mathbf{h}_{{\rm S}_1 {\rm D}_1}\|^2}{N_0}\right).
\end{flalign}
Therefore, the WSR maximization problem for fixed power scenario can be mathematically expressed as
\begin{equation*}
\begin{aligned}
\mathbf{P}_1:\,\,& \underset{\bm{\tau}, \bm{\omega}, \textcolor[rgb]{0.00,0.00,0.00}{R_{{\rm S}_1}, R_{{\rm S}_2}}}{\text{maximize}}
& & \alpha_1 R_{{\rm S}_1} + \alpha_2 R_{{\rm S}_2} \\
& \text{subject to}
& & (\ref{eqn:tau}), (\ref{eqn:wnorm}), (\ref{eqn:renergy}), (\ref{eqn:r12}), (\ref{eqn:rs2fp}), (\ref{eqn:rs1fp}) .
\end{aligned}
\end{equation*}

\textcolor[rgb]{0.00,0.00,0.00}{It is worth nothing that Problem $\mathbf{P}_1$ can be regarded as a general form of the data rate maximization oriented design for the considered cooperative WPCN. Particularly, when $\alpha_1=\alpha_2\neq 0$, the problem turns to be a rate-constrained sum-rate maximization. When $\alpha_i=0$ and $\alpha_j\neq 0$, where $i,j\in\{1,2\}$ and $i\neq j$, the problem turns to be an optimization problem which maximizes the data rate of group $j$ while guaranteing the minimal required data rate of group $i$.}
Nevertheless, it is observed that the right sides of (\ref{eqn:rs2fp}) and (\ref{eqn:rs1fp}) are non-linear w.r.t. $\bm{\tau}$ and $\bm{\omega}$, so constraints (\ref{eqn:rs2fp}) and (\ref{eqn:rs1fp}) are non-convex sets. Moreover, (\ref{eqn:renergy}) and (\ref{eqn:r12}) are also non-convex sets w.r.t. $\bm{\tau}$ and $\bm{\omega}$. Therefore, $\mathbf{P}_1$ is not a convex problem and cannot be solved with known solution methods. Thus, we solve it as follows.

\subsubsection{Problem Transformation and Solution}
We observe that $\bm{\omega}$ always appears in a quadratic form as shown in constraints (\ref{eqn:wnorm}), (\ref{eqn:renergy}) and (\ref{eqn:rs1}). By defining $\bm{\Omega} \triangleq \bm{\omega} \bm{\omega}^H$, the three constraints (\ref{eqn:wnorm}), (\ref{eqn:renergy}) and (\ref{eqn:rs1}) can be re-interpreted as
\begin{equation}\label{eqn:wmat}
\text{Tr}(\bm{\Omega}) \leq 1,
\end{equation}
\begin{equation}\label{eqn:wrenergy}
\textcolor[rgb]{0.00,0.00,0.00}{\tau_4} P_{\rm{R}} \leq \eta \tau_1 P_{\rm{S}_1} \mathbf{h}_{\rm{S}_1 \rm{R}}^H \bm{\Omega} \mathbf{h}_{\rm{S}_1 \rm{R}}+ \eta \tau_2 P_{\rm{S}_1}| \mathbf{h}_{\rm{S}_1 \rm{R}}^H \tfrac{\mathbf{h}_{\rm{S}_1 \rm{D}_1}}{\parallel\mathbf{h}_{\rm{S}_1 \rm{D}_1}\parallel}|^2,
\end{equation}
and
\begin{equation}\label{eqn:wrs1}
R_{\rm{S}_1} \textcolor[rgb]{0.00,0.00,0.00}{\leq} \tau_1\mathcal{C}\left( \frac{P_{\rm{S}_1} \mathbf{h}_{\rm{S}_1 \rm{D}_1}^H \bm{\Omega} \mathbf{h}_{\rm{S}_1 \rm{D}_1}}{N_0}\right)
+ \textcolor[rgb]{0.00,0.00,0.00}{\tau_2}\mathcal{C}\left( \frac{P_{\rm{S}_1} \|\mathbf{h}_{\rm{S}_1 \rm{D}_1}\|^2}{N_0}\right).
\end{equation}

Note that in order to ensure that $\bm{\omega}$ could be recovered by $\bm{\Omega}$ uniquely, it must satisfy that
\begin{equation}\label{eqn:wsd}
\bm{\Omega} \succeq 0,
\end{equation}
and
\begin{equation}\label{eqn:wrank}
\text{rank}(\bm{\Omega}) = 1 .
\end{equation}
Therefore, by replacing $\bm{\omega}$ with $\bm{\Omega}$, problem $\mathbf{P}_1$ is equivalently transformed into the following problem $\mathbf{P}_1^{\prime}$,
\begin{equation*}
\begin{aligned}
\mathbf{P}_1^{\prime}:\,\,& \underset{\bm{\tau}, \bm{\Omega}, \textcolor[rgb]{0.00,0.00,0.00}{R_{{\rm S}_1}, R_{{\rm S}_2}}}{\text{maximize}}
& & \alpha_1 R_{\rm{S}_1} + \alpha_2 R_{\rm{S}_2} \\
& \text{subject to}
& & (\ref{eqn:tau}), (\ref{eqn:r12}), (\ref{eqn:rs2fp}), (\ref{eqn:wmat}), (\ref{eqn:wrenergy}), (\ref{eqn:wrs1}), (\ref{eqn:wsd}), (\ref{eqn:wrank}) .
\end{aligned}
\end{equation*}
Problem $\mathbf{P}_1^{\prime}$ is still not jointly convex w.r.t. $\bm{\tau}$ and $\bm{\Omega}$ even though the rank-one constraint (\ref{eqn:wrank}) is removed. However, it can be observed that when the rank-one constraint is dropped, for a given $\bm{\tau}$, it is convex w.r.t. $\bm{\Omega}$. Meanwhile, for a given $\bm{\Omega}$, it is convex w.r.t. $\bm{\tau}$. Therefore, the relaxed problem of $\mathbf{P}_1^{\prime}$ can be solved by using traditional alternative iteration solution method. \textcolor[rgb]{0.00,0.00,0.00}{Nevertheless, with the traditional solution method, the convergence of the iteration can be proved, but it cannot be theoretically proved that the global optimal solution can always be guaranteed.} \textcolor[rgb]{0.00,0.00,0.00}{Instead,} we design a new solution method as follows, which is capable of finding the global optimal solution for Problem $\mathbf{P}_1^{\prime}$.

Define a new matrix variable $\bm{\digamma} \in \mathbb{C}^{N\times N}$ such that $\bm{\digamma} = \tau_1 \bm{\Omega}$. According to (\ref{eqn:wsd}) and (\ref{eqn:wrank}), it is known that
\begin{equation}\label{eqn:vsd}
\bm{\digamma} \succeq 0,
\end{equation}
and
\begin{equation}\label{eqn:vrank}
\text{rank}(\bm{\digamma}) = 1 .
\end{equation}
By substitution of $\bm{\Omega}=\frac{\bm{\digamma}}{\tau_1}$ into (\ref{eqn:wmat}) and (\ref{eqn:wrs1}), the two constraints (\ref{eqn:wmat}) and (\ref{eqn:wrs1}) can be respectively re-expressed by
\begin{equation}\label{eqn:vmat}
  \text{Tr}(\bm{\digamma}) \leq \tau_1,
\end{equation}
and
\begin{equation}\label{eqn:vrs1}
  R_{\rm{S}_1} \textcolor[rgb]{0.00,0.00,0.00}{\leq} \tau_1\mathcal{C}\left( \frac{ P_{\rm{S}_1} \text{Tr} (\bm{\digamma} \mathbf{h}_{\rm{S}_1 \rm{D}_1} \mathbf{h}_{\rm{S}_1 \rm{D}_1}^H )}{N_0 \tau_1}\right) + \textcolor[rgb]{0.00,0.00,0.00}{\tau_2}\mathcal{C}\left( \frac{P_{\rm{S}_1}  \|\mathbf{h}_{\rm{S}_1 \rm{D}_1}\|^2}{N_0}\right)
\end{equation}
.

Moreover, let $\phi_4 = \textcolor[rgb]{0.00,0.00,0.00}{\tau_4} P_{\rm{R}}$. (\ref{eqn:wrenergy}) and (\ref{eqn:rs2fp}) can be respectively rewritten as
\begin{equation}\label{eqn:vrenergy}
  \phi_4 \leq \eta \textcolor[rgb]{0.00,0.00,0.00}{P_{\rm{S}_1}} \text{Tr} (\bm{\digamma} \mathbf{h}_{\rm{S}_1 \rm{R}} \mathbf{h}_{\rm{S}_1 \rm{R}}^H )+ \textcolor[rgb]{0.00,0.00,0.00}{\eta \tau_2 P_{\rm{S}_1}| \mathbf{h}_{\rm{S}_1 \rm{R}}^H \tfrac{\mathbf{h}_{\rm{S}_1 \rm{D}_1}}{\parallel\mathbf{h}_{\rm{S}_1 \rm{D}_1}\parallel}|^2}
\end{equation}
and
%\begin{flalign}\label{eqn:vrs22}
%  R_{\rm{S}_2} = \min \Bigg\{&\textcolor[rgb]{0.00,0.00,0.00}{\tau_3}\mathcal{C}\left( \frac{P_{\rm{S}_2}|h_{\rm{S}_2 \rm{R}}|^2}{N_0}\right),\\
%&\textcolor[rgb]{0.00,0.00,0.00}{\tau_3}\mathcal{C}\left( \frac{P_{\rm{S}_2}|h_{\rm{S}_2 \rm{D}_2}|^2}{N_0}\right)+\textcolor[rgb]{0.00,0.00,0.00}{\tau_4}\mathcal{C}\left( \frac{\phi_4|h_{\rm{R} \rm{D}_2}|^2}{N_0 \textcolor[rgb]{0.00,0.00,0.00}{\tau_4}}\right)\Bigg\}.\nonumber
%\end{flalign}
\begin{flalign}\label{eqn:vrs22}
R_{\rm{S}_2} \textcolor[rgb]{0.00,0.00,0.00}{\leq} \min \Bigg\{&\textcolor[rgb]{0.00,0.00,0.00}{\tau_3}\mathcal{C}\left( \frac{P_{\rm{S}_2}|h_{\rm{S}_2 \rm{R}}|^2}{N_0}\right),\\
&\textcolor[rgb]{0.00,0.00,0.00}{\tau_3}\mathcal{C}\left( \frac{P_{\rm{S}_2}|h_{\rm{S}_2 \rm{D}_2}|^2}{N_0}\right)+\textcolor[rgb]{0.00,0.00,0.00}{\tau_4}\mathcal{C}\left( \frac{\phi_4|h_{\rm{R} \rm{D}_2}|^2}{N_0 \textcolor[rgb]{0.00,0.00,0.00}{\tau_4}}\right)\Bigg\}.\nonumber
\end{flalign}

With above variable substitution operations, i.e., $\bm{\digamma} = \tau_1 \bm{\Omega}$, and $\phi_4 = \textcolor[rgb]{0.00,0.00,0.00}{\tau_4} P_{\rm{R}}$, Problem $\mathbf{P}_1^{\prime}$ is equivalently transformed into the following Problem $\mathbf{P}_1^{\prime\prime}$,
\begin{flalign*}
\mathbf{P}_1^{\prime\prime}:\,\, \underset{\bm{\tau}, \bm{\digamma}, \phi_4, \textcolor[rgb]{0.00,0.00,0.00}{R_{{\rm S}_1}, R_{{\rm S}_2}}}{\text{maximize}}\,\,
&\alpha_1 R_{\rm{S}_1} + \alpha_2 R_{\rm{S}_2} \nonumber\\
\text{subject to}\,\,
&(\ref{eqn:tau}), (\ref{eqn:r12}), (\ref{eqn:vsd}),\nonumber\\
&(\ref{eqn:vrank}), (\ref{eqn:vmat}), (\ref{eqn:vrs1}), (\ref{eqn:vrenergy}), (\ref{eqn:vrs22}).\nonumber
\end{flalign*}

By dropping the rank-1 constraint in (\ref{eqn:vrank}), we obtain that
\begin{equation*}
\begin{aligned}
\mathbf{P}_1^{\prime\prime\prime}:\,\,& \underset{\bm{\tau}, \bm{\digamma}, \phi_4, \textcolor[rgb]{0.00,0.00,0.00}{R_{{\rm S}_1}, R_{{\rm S}_2}}}{\text{minimize}}
& & - \alpha_1 R_{\rm{S}_1} - \alpha_2 R_{\rm{S}_2} \\
& \text{subject to}
& & (\ref{eqn:tau}), (\ref{eqn:r12}), (\ref{eqn:vsd}),(\ref{eqn:vmat}), (\ref{eqn:vrs1}), (\ref{eqn:vrenergy}), (\ref{eqn:vrs22}).
\end{aligned}
\end{equation*}
\newtheorem{prop}{Proposition}
\newtheorem{lemma}{Lemma}
\textcolor[rgb]{0.00,0.00,0.00}{\begin{prop}\label{pop1}
$\mathbf{P}_1^{\prime\prime\prime}$ is a convex problem.
\end{prop}
\begin{IEEEproof}
The objective function of Problem $\mathbf{P}_1^{\prime\prime\prime}$ is linear. The constraints (\ref{eqn:tau}), (\ref{eqn:r12}), (\ref{eqn:vsd}), (\ref{eqn:vmat}) and (\ref{eqn:vrenergy}) are all convex sets. Moreover, as $y\log(1+\frac{x}{y})$ is a perspective
function of concave function $\log(1+x)$, which is joint concave w.r.t $x$ and $y$\cite{Boyd}, it can be proved that (\ref{eqn:vrs1}) and (\ref{eqn:vrs22}) are also convex sets. Thus, we arrive at Proposition \ref{pop1}.
\end{IEEEproof}}
Via the relaxation described above, the non-convex Problem $\mathbf{P}_1^{\prime\prime}$ is transformed to be the convex Problem of $\mathbf{P}_1^{\prime\prime\prime}$ by using the SDR \cite{Luo}. Therefore, by employing some known solution methods, e.g., interior point method, for convex problems \cite{Boyd}, the optimal $[\bm{\tau}^*,\bm{\digamma}^*, \phi_4^*]$ of Problem $\mathbf{P}_1^{\prime\prime\prime}$ can be obtained.

\subsubsection{Global Optimum Analysis for Our Proposed Solution Method}
\textcolor[rgb]{0.00,0.00,0.00}{Note that our goal is to find the optimal $[{\bm \tau}^*, {\bm \omega}^*]$ for Problem $\mathbf{P}_1$ rather than the optimal $[\bm{\tau}^*,\bm{\digamma}^*, \bm{\phi}^*]$.} It is known that, only when $\text{rank}(\bm{\digamma}^*) = 1$, $[\bm{\tau}^*,\bm{\digamma}^*, \phi_4^*]$ is also the optimal solution of Problem $\mathbf{P}_1^{\prime\prime}$. In this case, the optimal  $[\bm{\tau}^*,\bm{\omega}^*]$ can be derived accordingly. Therefore, the key question lies in the rank of $\bm{\digamma}^*$. Fortunately, we found that there exists an optimal $\bm{\digamma}^*$ such that $\text{rank}(\bm{\digamma}^*) = 1$ for Problem $\mathbf{P}_1^{\prime\prime\prime}$, which means the global optimum of the primary Problem $\mathbf{P}_1$ can be guaranteed.

Now we analyse the rank of $\bm{\digamma}^*$ with Theorem \ref{theo:vrank1fp}. Before that, we present Lemma \ref{lem:vrank1}, which was proved in \cite{Huang}, for emphasis as follows.

\begin{lemma}\cite{Huang}\label{lem:vrank1}
Consider a problem $\mathbf{P}_0$,
\begin{equation*}
\begin{aligned}
\mathbf{P}_0:\,\,& \underset{\mathbf{X}_1, \ldots, \mathbf{X}_L}{\text{minimize}}
& & \sum\nolimits_{l = 1}^L \text{Tr}(\mathbf{C}_l \mathbf{X}_l)\\
& \text{subject to}
& & \sum\nolimits_{l = 1}^{L} \text{Tr}(\mathbf{A}_{ml} \mathbf{X}_l) \unrhd_m b_m, \,\,m = 1,\ldots,M,\\
&&& \mathbf{X}_l \succeq 0, \,\,l = 1,\ldots,L,
\end{aligned}
\end{equation*}
where $\mathbf{C}_l,l = 1,\ldots,L$ and $\mathbf{A}_{ml},m = 1,\ldots,M,l = 1,\ldots,L$ are Hermitian matrices, $b \in \mathbb{R}$, $\unrhd_m \in \{\geq, =, \leq\},m = 1,\ldots,M$ and the variables $\mathbf{X}_l,l = 1,\ldots,L$ are Hermitian matrices.
If Problem $\mathbf{P}_0$ and its dual are solvable, then the Problem $\mathbf{P}_0$ has always an optimal solution $(\mathbf{X}_1^*, \ldots, \mathbf{X}_L^*)$ such that
$
\sum\nolimits_{l = 1}^{L} \text{rank}^2 (\mathbf{X}_l^*) \leq M.
$
\end{lemma}

\newtheorem{theorem}{Theorem}
\begin{theorem}\label{theo:vrank1fp}
There exists an optimal $\bm{\digamma}^*$ of Problem $\mathbf{P}_1^{\prime\prime\prime}$ such that $\text{rank}(\bm{\digamma}^*) = 1$.
\end{theorem}
\begin{IEEEproof}
The proof can be found in Appendix \ref{App:T1}.
\end{IEEEproof}
\newtheorem{corollary}{Corollary}
\begin{corollary}\label{rmk:p1}
  The global optimal solution to Problem $\textbf{P}_1$ is guaranteed by using our proposed solution method.
\end{corollary}
\begin{IEEEproof}
 $\mathbf{P}_1$, $\mathbf{P}_1^{\prime}$ and $\mathbf{P}_1^{\prime\prime}$ are equivalent to each other. It is known that once the optimal solution of $\mathbf{P}_1^{\prime\prime\prime}$ satisfies the rank-one constraint, it is equivalent to $\mathbf{P}_1$, $\mathbf{P}_1^{\prime}$ and $\mathbf{P}_1^{\prime\prime}$. Theorem \ref{theo:vrank1fp} declares that
$\mathbf{P}_1^{\prime\prime\prime}$ has a rank-one optimal solution. Therefore, the optimal solution for Problem $\mathbf{P}_1$ can always be found by using our proposed solution method.
\end{IEEEproof}

\subsection{Flexible Power Scenario}

\subsubsection{Problem Formulation}
In flexible power scenario, $\rm{S}_1$ and $\rm{S}_2$ are allowed to transmit information/energy in different phases with different power, but the averaged power over each fading block is confined by $P_{\rm{S}_1}$ and $P_{\rm{S}_2}$ respectively. That is, the consumed powers at $\rm{S}_1$ and $\rm{S}_2$ respectively satisfy that
\begin{equation}\label{eqn:powers1}
\tau_1 P_{\rm{S}_1}^{(1)} + \textcolor[rgb]{0.00,0.00,0.00}{\tau_2} P_{\rm{S}_1}^{\textcolor[rgb]{0.00,0.00,0.00}{(2)}} \leq P_{\rm{S}_1},
\end{equation}
and
\begin{equation}\label{eqn:powers2}
\textcolor[rgb]{0.00,0.00,0.00}{\tau_3} P_{\rm{S}_2}^{\textcolor[rgb]{0.00,0.00,0.00}{(3)}} \leq P_{\rm{S}_2} .
\end{equation}

For clarity, we define $\mathbf{P} \triangleq [P_{\rm{S}_1}^{(1)}\,\,P_{\rm{S}_1}^{\textcolor[rgb]{0.00,0.00,0.00}{(2)}}\,\,P_{\rm{S}_2}^{\textcolor[rgb]{0.00,0.00,0.00}{(3)}}]^T$, \textcolor[rgb]{0.00,0.00,0.00}{which can be regarded as the power allocation vector for the four phases}. Thus, the WSR maximization problem
can be mathematically expressed by
\begin{equation*}
\begin{aligned}
\mathbf{P}_2:\,\,& \underset{\bm{\tau}, \bm{\omega}, \mathbf{P},\textcolor[rgb]{0.00,0.00,0.00}{R_{{\rm S}_1}, R_{{\rm S}_2}}}{\text{maximize}}
& & \alpha_1 R_{\rm{S}_1} + \alpha_2 R_{\rm{S}_2} \\
& \text{subject to}
& & (\ref{eqn:tau}), (\ref{eqn:wnorm}), (\ref{eqn:renergy}), (\ref{eqn:rs2}), (\ref{eqn:rs1}), (\ref{eqn:r12}), (\ref{eqn:powers1}), (\ref{eqn:powers2}).
\end{aligned}
\end{equation*}

Compared with Problem $\mathbf{P}_1$ for the fixed power scenario, in Problem $\mathbf{P}_2$, the power $\mathbf{P}$ consumed in each phase at the two sources are jointly optimized with $\bm{\tau}$ and $\bm{\omega}$. Similar to Problem $\mathbf{P}_1$, it can be observed that Problem $\mathbf{P}_2$ is also non-convex. So we solve it as follows.

\subsubsection{Problem Transformation and Solution}
Like the solution method designed for Problem $\mathbf{P}_1$, we also deal with Problem $\mathbf{P}_2$ by transforming it into a convex problem through variable substitution operations and SDR at first and then solve it efficiently.

We also use the definition of $\bm{\Omega} \triangleq \bm{\omega} \bm{\omega}^H$ by introducing a semi-definite square matrix $\bm{\Omega}\succeq 0$. Then, (\ref{eqn:renergy}) can be equivalently replaced by (\ref{eqn:wrenergy}), and (\ref{eqn:rs1}) can be re-expressed by
\begin{equation}\label{eqn:rs1ap}
  R_{\rm{S}_1} \textcolor[rgb]{0.00,0.00,0.00}{\leq} \tau_1\mathcal{C}\left(\frac{P_{\rm{S}_1}^{(1)}\mathbf{h}_{\rm{S}_1 \rm{D}_1}^H \bm{\Omega} \mathbf{h}_{\rm{S}_1 \rm{D}_1}}{N_0}\right) + \textcolor[rgb]{0.00,0.00,0.00}{\tau_2}\mathcal{C}\left(\frac{P_{\rm{S}_1}^{\textcolor[rgb]{0.00,0.00,0.00}{(2)}} \|\mathbf{h}_{\rm{S}_1 \rm{D}_1}\|^2}{N_0}\right).
\end{equation}

Consequently, with the rank-one constraint of $\bm{\Omega}$, i.e., $\text{rank}(\bm{\Omega}) = 1$, Problem $\mathbf{P}_2$ is equivalently transformed into the following Problem $\mathbf{P}_2^{\prime}$, i.e.,
\begin{flalign*}
\mathbf{P}_2^{\prime}:\,\,\underset{\bm{\tau}, \bm{\Omega}, \mathbf{P},\textcolor[rgb]{0.00,0.00,0.00}{R_{{\rm S}_1}, R_{{\rm S}_2}}}{\text{maximize}}\,\,
&\alpha_1 R_{\rm{S}_2} + \alpha_2 R_{\rm{S}_1} \\
\text{subject to}\,\,
& (\ref{eqn:tau}), (\ref{eqn:rs2}), (\ref{eqn:r12}), (\ref{eqn:wmat}), (\ref{eqn:wrenergy}),\\
& (\ref{eqn:wsd}),
(\ref{eqn:wrank}), (\ref{eqn:powers1}), (\ref{eqn:powers2}),  (\ref{eqn:rs1ap}).
\end{flalign*}

Since Problem $\mathbf{P}_2^{\prime}$ is still non-convex, we further adopt the following variable substitutions by introducing five new variables, i.e.,
\begin{equation}\label{eqn:trans}
  \begin{cases}
    \phi_1 = & \tau_1 P_{\rm{S}_1}^{(1)},\,\,
    \phi_2 =  \textcolor[rgb]{0.00,0.00,0.00}{\tau_2}P_{\rm{S}_1}^{\textcolor[rgb]{0.00,0.00,0.00}{(2)}},\\
    \phi_3 = &\textcolor[rgb]{0.00,0.00,0.00}{\tau_3} P_{\rm{S}_2}^{\textcolor[rgb]{0.00,0.00,0.00}{(3)}},\,\,
    \phi_4 = \textcolor[rgb]{0.00,0.00,0.00}{\tau_4} P_{\rm{R}},\,\,
\\
    \bm{G} = & \tau_1 P_{\rm{S}_1}^{(1)}\bm{\Omega}=\phi_1\bm{\Omega},
  \end{cases}
\end{equation}
\textcolor[rgb]{0.00,0.00,0.00}{with
\begin{equation}\label{eqn:gsd}
\bm{G}\succeq 0
\end{equation}
and
\begin{equation}\label{eqn:rankG}
\textrm{rank}(\bm{G})=1.
\end{equation}}

With these linear definitions, (\ref{eqn:wmat}), (\ref{eqn:wrenergy}), (\ref{eqn:wsd}) and (\ref{eqn:wrank}) can be respectively replaced by (\ref{eqn:vmat}), (\ref{eqn:vrenergy}), (\ref{eqn:gsd}) and (\ref{eqn:rankG}). Moreover, (\ref{eqn:rs2}), (\ref{eqn:powers1}), (\ref{eqn:powers2}), (\ref{eqn:vmat}) and (\ref{eqn:rs1ap}) are respectively transformed into
%\begin{flalign}\label{eqn:rs2ap}
%  R_{\rm{S}_2}=\min &\left\{\textcolor[rgb]{0.00,0.00,0.00}{\tau_3}\mathcal{C}\Bigg(\frac{\phi_3 |h_{\rm{S}_2 R}|^2}{N_0 \textcolor[rgb]{0.00,0.00,0.00}{\tau_3}}\right),\\
%&\quad\,\,\textcolor[rgb]{0.00,0.00,0.00}{\tau_3}\mathcal{C}\left(\frac{\phi_3 |h_{\rm{S}_2 {\rm D}_2}|^2}{N_0 \textcolor[rgb]{0.00,0.00,0.00}{\tau_3}}\right)+\textcolor[rgb]{0.00,0.00,0.00}{\tau_4}\mathcal{C}\left(\frac{\phi_4|h_{{\rm R} {\rm D}_2}|^2}{N_0 \textcolor[rgb]{0.00,0.00,0.00}{\tau_4}}\right)\Bigg\},\nonumber
%\end{flalign}
\begin{flalign}\label{eqn:rs2ap}
  R_{\rm{S}_2} \textcolor[rgb]{0.00,0.00,0.00}{\leq} \min \Bigg\{&\textcolor[rgb]{0.00,0.00,0.00}{\tau_3}\mathcal{C}\Big(\frac{\phi_3 |h_{\rm{S}_2 R}|^2}{N_0 \textcolor[rgb]{0.00,0.00,0.00}{\tau_3}}\Big),\\
&\textcolor[rgb]{0.00,0.00,0.00}{\tau_3}\mathcal{C}\left(\frac{\phi_3 |h_{\rm{S}_2 {\rm D}_2}|^2}{N_0 \textcolor[rgb]{0.00,0.00,0.00}{\tau_3}}\right)+\textcolor[rgb]{0.00,0.00,0.00}{\tau_4}\mathcal{C}\left(\frac{\phi_4|h_{{\rm R} {\rm D}_2}|^2}{N_0 \textcolor[rgb]{0.00,0.00,0.00}{\tau_4}}\right)\Bigg\},\nonumber
\end{flalign}
\begin{equation}\label{eqn:powers1ap}
  \phi_1 + \phi_2 \leq P_{\rm{S}_1} ,
\end{equation}
\begin{equation}\label{eqn:powers2ap}
  \phi_3 \leq P_{\rm{S}_2} ,
\end{equation}
\begin{equation}\label{eqn:vmatap}
\textcolor[rgb]{0.00,0.00,0.00}{ \text{Tr}(\bm{G}) \leq \phi_1}
\end{equation}
and
\begin{equation}\label{eqn:rs1ap2}
  R_{\rm{S}_1} \textcolor[rgb]{0.00,0.00,0.00}{\leq} \tau_1\mathcal{C}\left(\frac{\text{Tr} (\bm{G} \mathbf{h}_{\rm{S}_1 \rm{D}_1} \mathbf{h}_{\rm{S}_1 \rm{D}_1}^H )}{N_0 \tau_1}\right) + \textcolor[rgb]{0.00,0.00,0.00}{\tau_2}\mathcal{C}\left( \frac{\phi_2 \|\mathbf{h}_{\rm{S}_1 \rm{D}_1}\|^2}{N_0 \textcolor[rgb]{0.00,0.00,0.00}{\tau_2}}\right) .
\end{equation}

Let $\bm{\phi}=[\phi_1\,\,\phi_2\,\,\phi_3\,\,\phi_4]^T$. With the definitions in (\ref{eqn:trans}),  Problem $\mathbf{P}_2^{\prime}$ can be equivalently transformed into the following Problem $\mathbf{P}_2^{\prime\prime}$,
%\begin{equation}
%\begin{aligned}
%\mathbf{P}_1^{\prime}:\,\,& \underset{\bm{\tau}, \bm{\Omega}}{\text{maximize}}
%& & \alpha_1 R_{\rm{S}_2} + \alpha_2 R_{\rm{S}_1} \\
%& \text{subject to}
%& & \mathbf{1}^T \bm{\tau} = 1, \\
%&&& \bm{\tau} \succeq \mathbf{0}, \\
%&&& \text{Tr}(\bm{\Omega}) \leq P_{\rm{S}_1}^1 , \\
%&&& \tau_1 P_{\rm{S}_1}^1 + \textcolor[rgb]{0.00,0.00,0.00}{\tau_2} P_{\rm{S}_1}^4 \leq P_{\rm{S}_1} , \\
%&&& \textcolor[rgb]{0.00,0.00,0.00}{\tau_3} P_{\rm{S}_2}^2 \leq P_{\rm{S}_2} , \\
%&&& \textcolor[rgb]{0.00,0.00,0.00}{\tau_4} P_{\rm{R}} \leq \eta \tau_1 \mathbf{h}_{\rm{S}_1 \rm{R}}^H \bm{\Omega} \mathbf{h}_{\rm{S}_1 \rm{R}}, \\
%&&& \bm{\Omega} \succeq 0, \\
%&&& \text{rank}(\bm{\Omega}) = 1 , \\
%&&& R_{\rm{S}_2} \geq r_{\rm{S}_2} , \\
%&&& R_{\rm{S}_1} \geq r_{\rm{S}_1} , \\
%& \text{where}
%& & R_{\rm{S}_1} = \tau_1\log\left(1 + \frac{\mathbf{h}_{\rm{S}_1 \rm{D}_1}^H \bm{\Omega} \mathbf{h}_{\rm{S}_1 \rm{D}_1}}{N_0}\right)
%+ \textcolor[rgb]{0.00,0.00,0.00}{\tau_2}\log\left(1 + \frac{P_{\rm{S}_1}^4 \|\mathbf{h}_{\rm{S}_1 \rm{D}_1}\|^2}{N_0}\right).
%\end{aligned}
%\end{equation}

\begin{flalign*}
\mathbf{P}_2^{\prime\prime}:\,\, \underset{\bm{\tau}, \bm{\textcolor[rgb]{0.00,0.00,0.00}{G}}, \bm{\phi},\textcolor[rgb]{0.00,0.00,0.00}{R_{{\rm S}_1}, R_{{\rm S}_2}}}{\text{maximize}}
& \alpha_1 R_{\rm{S}_2} + \alpha_2 R_{\rm{S}_1} \\
\text{subject to}\,\,
& (\ref{eqn:tau}),(\ref{eqn:r12}), (\ref{eqn:vrenergy}), (\ref{eqn:gsd}), (\ref{eqn:rankG}),\\
&(\ref{eqn:rs2ap}), (\ref{eqn:powers1ap}), (\ref{eqn:powers2ap}), (\ref{eqn:vmatap}), (\ref{eqn:rs1ap2}).
\end{flalign*}

It can be seen that the objective function of Problem $\mathbf{P}_2^{\prime\prime}$ is concave and all constraints except (\ref{eqn:rankG}) are convex sets. Therefore, by using SDR method with the dropping of (\ref{eqn:rankG}), Problem $\mathbf{P}_2^{\prime\prime}$ can be relaxed to a convex problem as follows,
\begin{flalign*}
\mathbf{P}_2^{\prime\prime\prime}:\,\, \underset{\bm{\tau}, \bm{\textcolor[rgb]{0.00,0.00,0.00}{G}}, \bm{\phi}, \textcolor[rgb]{0.00,0.00,0.00}{R_{{\rm S}_1}, R_{{\rm S}_2}}}{\text{minimize}}\,\,
& - \alpha_1 R_{\rm{S}_1} - \alpha_2 R_{\rm{S}_2} \\
\text{subject to}\,\,
& (\ref{eqn:tau}), (\ref{eqn:r12}), (\ref{eqn:vrenergy}), (\ref{eqn:gsd}),\\
&(\ref{eqn:rs2ap}), (\ref{eqn:powers1ap}), (\ref{eqn:powers2ap}), (\ref{eqn:vmatap}), (\ref{eqn:rs1ap2}).
\end{flalign*}
\textcolor[rgb]{0.00,0.00,0.00}{\begin{prop}\label{pop2}
$\mathbf{P}_2^{\prime\prime\prime}$ is a convex problem.
\end{prop}
\begin{IEEEproof}
 The proof is similar to that of Proposition \ref{pop2}, which is omitted here.
\end{IEEEproof}}
Therefore, the optimal solution $[\bm{\tau}^*,\bm{\textcolor[rgb]{0.00,0.00,0.00}{G}}^*, \bm{\phi}^*]$ of Problem $\mathbf{P}_2^{\prime\prime\prime}$ can be obtained by using some known solution methods.

\subsubsection{Global Optimum Analysis for Our Proposed Solution Method}
Similar to the situation of Problem $\mathbf{P}_1^{\prime\prime\prime}$, only when $\text{rank}(\bm{\textcolor[rgb]{0.00,0.00,0.00}{G}}^*) = 1$, $[\bm{\tau}^*,\bm{\textcolor[rgb]{0.00,0.00,0.00}{G}}^*, \bm{\phi}^*]$ is also the optimal solution of Problem $\mathbf{P}_2^{\prime\prime}$. In this case, the optimal  $[\bm{\tau}^*,\bm{\omega}^*,\mathbf{P}^*]$ can be derived accordingly. Therefore, the key question lies in the rank of $\bm{\textcolor[rgb]{0.00,0.00,0.00}{G}}^*$. Fortunately, we also found that $\text{rank}(\bm{\textcolor[rgb]{0.00,0.00,0.00}{G}}^*) = 1$ always holds for Problem $\mathbf{P}_2^{\prime\prime\prime}$, which means the global optimum of the primary Problem $\mathbf{P}_2$ also can be guaranteed by our adopted variable substitutions and SDR.

Now we analyse the rank of $\bm{\textcolor[rgb]{0.00,0.00,0.00}{G}}^*$ for the average power constrained scenario with Theorem \ref{theo:vrank1ap}.

\begin{theorem}\label{theo:vrank1ap}
There exists an optimal $\bm{\textcolor[rgb]{0.00,0.00,0.00}{G}}^*$ of Problem $\mathbf{P}_2^{\prime\prime\prime}$ such that $\text{rank}(\bm{\textcolor[rgb]{0.00,0.00,0.00}{G}}^*) = 1$.
\end{theorem}
\begin{IEEEproof}
The proof can be found in Appendix \ref{App:T2}.
\end{IEEEproof}

\begin{corollary}\label{rmk:p2}
The optimal solution of Problem $\mathbf{P}_2$ to the flexible power scenario is guaranteed by using our proposed method.
\end{corollary}
\begin{IEEEproof} The proof of Corollary 2 is similar to that of Corollary 1. $\mathbf{P}_2$, $\mathbf{P}_2^{\prime}$ and $\mathbf{P}_2^{\prime\prime}$ are equivalent to each other. Theorem \ref{theo:vrank1ap} declares that
$\mathbf{P}_2^{\prime\prime\prime}$ has a rank-one optimal solution. Therefore, the optimal solution to Problem $\mathbf{P}_2$ can always be found by using our proposed solution method.
\end{IEEEproof}

\section{Power-minimization Design}
Besides the throughput maximization design, the energy-saving design is another essential objective for practical energy-constrained wireless networks, e.g., WSNs, WPANs and WBANs, to extend their life time. Therefore, in this section, we investigate the minimum energy consumption design for the considered cooperative WPCN described in Section II. Our goal is to jointly optimize the beamforming, time allocation and power allocation to minimize the system total consumed power while guaranteeing the required information rates of the two groups.

\subsection{Problem Formulation}
As described in Section III, $\rm{S}_1$ transmits signals in the \textcolor[rgb]{0.00,0.00,0.00}{first and the second phases}, while $\rm{S}_2$ transmits signals only \textcolor[rgb]{0.00,0.00,0.00}{in the third phase}. Specifically, in the first phase, the consumed energy at $\rm{S}_1$ is $\tau_1 P_{\rm{S}_1}^{(1)}\| \bm{\omega} \|^2$, where $\| \bm{\omega} \|^2\leq1$. In the \textcolor[rgb]{0.00,0.00,0.00}{second} phase, the consumed energy at $\rm{S}_1$ is $\textcolor[rgb]{0.00,0.00,0.00}{\tau_2} P_{\rm{S}_1}^{\textcolor[rgb]{0.00,0.00,0.00}{(2)}}$. In the \textcolor[rgb]{0.00,0.00,0.00}{third} phase, the consumed energy at $\rm{S}_2$ is $\textcolor[rgb]{0.00,0.00,0.00}{\tau_3} P_{\rm{S}_2}^{\textcolor[rgb]{0.00,0.00,0.00}{(3)}}$. As a result, the total consumed energy for information trasnmission is $\tau_1 \| \bm{\omega} \|^2 + \textcolor[rgb]{0.00,0.00,0.00}{\tau_2} P_{\rm{S}_1}^{\textcolor[rgb]{0.00,0.00,0.00}{(2)}} + \textcolor[rgb]{0.00,0.00,0.00}{\tau_3} P_{\rm{S}_2}^{\textcolor[rgb]{0.00,0.00,0.00}{(3)}}$. Since the time period $T$ of the fading block is normalized to be 1, the total consumed power for the transmissions in the fading block is also expressed by
\begin{equation}\label{eqn:pavg}
  P_{\rm avg} = \tau_1 P_{\rm{S}_1}^{(1)}\| \bm{\omega} \|^2 + \textcolor[rgb]{0.00,0.00,0.00}{\tau_2} P_{\rm{S}_1}^{\textcolor[rgb]{0.00,0.00,0.00}{(2)}} + \textcolor[rgb]{0.00,0.00,0.00}{\tau_3} P_{\rm{S}_2}^{\textcolor[rgb]{0.00,0.00,0.00}{(3)}}.
\end{equation}

Therefore, the total power minimization problem under the minimal required data rates can be formulated as
\begin{equation*}
\begin{aligned}
\mathbf{P}_3:\,\,& \underset{\bm{\tau}, \bm{\omega}, \mathbf{P}}{\text{minimize}}
& & \textcolor[rgb]{0.00,0.00,0.00}{\tau_3} P_{\rm{S}_2}^{\textcolor[rgb]{0.00,0.00,0.00}{(3)}} + \tau_1 P_{\rm{S}_1}^{(1)}\| \bm{\omega} \|^2 + \textcolor[rgb]{0.00,0.00,0.00}{\tau_2} P_{\rm{S}_1}^{\textcolor[rgb]{0.00,0.00,0.00}{(2)}} \\
& \text{subject to}
& & (\ref{eqn:tau}), (\ref{eqn:wnorm}), (\ref{eqn:rs2}), (\ref{eqn:renergy}), (\ref{eqn:rs1}), (\ref{eqn:r12}),
\end{aligned}
\end{equation*}
which is also not jointly convex w.r.t. $\bm{\tau}$, $\bm{\omega}$ and \textbf{P} due to constraints (\ref{eqn:renergy}) and (\ref{eqn:r12}), so that it cannot be solved directly by using known solution methods. Therefore, we solve it as follows.

\subsection{Problem Transformation and Solution}
If we use the same variable substitution, i.e., $\bm{\Omega} = \bm{\omega}\bm{\omega}^H$ as described in Section III, constraints (\ref{eqn:renergy}) and (\ref{eqn:rs1}) of Problem $\mathbf{P}_3$ also can be equally replaced by (\ref{eqn:wrenergy}) and (\ref{eqn:rs1ap}), respectively. And, its objective function in (\ref{eqn:pavg}) can be rewritten as
\begin{equation}\label{eqn:wpavg}
  P_{\rm avg} = \textcolor[rgb]{0.00,0.00,0.00}{\tau_3} P_{\rm{S}_2}^{\textcolor[rgb]{0.00,0.00,0.00}{(3)}} + \tau_1P_{\rm{S}_1}^{(1)}\text{Tr} (\bm{\Omega}) + \textcolor[rgb]{0.00,0.00,0.00}{\tau_2} P_{\rm{S}_1}^{\textcolor[rgb]{0.00,0.00,0.00}{(2)}}.
\end{equation}

In order to equivalently transform Problem $\mathbf{P}_3$ into the following Problem $\mathbf{P}_3^{\prime}$, $\bm{\Omega}$ must be semi-definite and rank one, as expressed by the constraints (\ref{eqn:wsd}) and (\ref{eqn:wrank}). Problem $\mathbf{P}_3^{\prime}$ then can be given by
\begin{equation*}
\begin{aligned}
\mathbf{P}_3^{\prime}:\,\,& \underset{\bm{\tau}, \bm{\Omega}, \mathbf{P}}{\text{minimize}}
& & \textcolor[rgb]{0.00,0.00,0.00}{\tau_3} P_{\rm{S}_2}^{\textcolor[rgb]{0.00,0.00,0.00}{(3)}} + \tau_1P_{\rm{S}_1}^{(1)}\text{Tr} (\bm{\Omega}) + \textcolor[rgb]{0.00,0.00,0.00}{\tau_2} P_{\rm{S}_1}^{\textcolor[rgb]{0.00,0.00,0.00}{(2)}} \\
& \text{subject to}
& & (\ref{eqn:tau}), (\ref{eqn:rs2}), (\ref{eqn:r12}), (\ref{eqn:wmat}),  (\ref{eqn:wrenergy}),
 (\ref{eqn:wsd}), (\ref{eqn:wrank}),  (\ref{eqn:rs1ap}),
\end{aligned}
\end{equation*}
which is an equivalent transformation of Problem $\mathbf{P}_3$.
Since Problem $\mathbf{P}_3^{\prime}$ is still non-convex, we further transform it to be the following Problem $\mathbf{P}_3^{\prime\prime}$ by using the variable substitution defined in (\ref{eqn:trans}).
%\begin{equation}\label{eqn:trans}
%  \begin{cases}
%    \bm{\digamma} = & \tau_1 \bm{\Omega},\\
%    \phi_4 = & \textcolor[rgb]{0.00,0.00,0.00}{\tau_4} P_{\rm{R}}, \\
%    \phi_3 = & \textcolor[rgb]{0.00,0.00,0.00}{\tau_3} P_{\rm{S}_2}^{\textcolor[rgb]{0.00,0.00,0.00}{(3)}}, \\
%    \phi_1 = & \tau_1 P_{\rm{S}_1}^{(1)}, \\
%    \phi_2 = & \textcolor[rgb]{0.00,0.00,0.00}{\tau_2} P_{\rm{S}_1}^{\textcolor[rgb]{0.00,0.00,0.00}{(2)}}, \\
%  \end{cases}
%\end{equation}
%which is similar to (\ref{eqn:trans}).

To make Problem $\mathbf{P}_3^{\prime\prime}$ be an equivalent version of Problem $\mathbf{P}_3^{\prime}$, $\bm{G}$ also should satisfy the semi-definite constraint and rank-one constraint, which can be expressed by (\ref{eqn:gsd}) and (\ref{eqn:rankG}). Moreover, with (\ref{eqn:trans}), constraints (\ref{eqn:wrenergy}), (\ref{eqn:rs2}) and (\ref{eqn:rs1ap}) are replaced with (\ref{eqn:vrenergy}), (\ref{eqn:rs2ap}) and (\ref{eqn:rs1ap2}) respectively. The objective function (\ref{eqn:wpavg}) of Problem $\mathbf{P}_3^{\prime}$ can be transformed into
%\begin{equation}\label{eqn:vpavg}
$P_{\rm avg} = \phi_3 + \text{Tr} (\bm{G}) + \phi_2$.
%\end{equation}
Also, let $\bm{\phi}=[\phi_4\,\,\phi_3\,\,\phi_1\,\,\phi_2]^T$. Problem $\mathbf{P}_3^{\prime\prime}$ can be given by
\begin{equation*}
\begin{aligned}
\mathbf{P}_3^{\prime\prime}:\,\,& \underset{\bm{\tau}, \bm{G}, \bm{\phi}}{\text{minimize}}
& & \phi_3 + \text{Tr} (\bm{G}) + \phi_2 \\
& \text{subject to}
& & (\ref{eqn:tau}), (\ref{eqn:r12}), (\ref{eqn:gsd}), (\ref{eqn:rankG}), (\ref{eqn:vmatap}), (\ref{eqn:vrenergy}), (\ref{eqn:rs2ap}), (\ref{eqn:rs1ap2}).
\end{aligned}
\end{equation*}

It can be seen that the objective function of Problem $\mathbf{P}_3^{\prime\prime}$ is convex and all constraints except the rank-one constraint (\ref{eqn:rankG}) are convex sets. Therefore, by using SDR method with the dropping of (\ref{eqn:rankG}), Problem $\mathbf{P}_3^{\prime\prime}$ can be relaxed to a convex problem as follows,
\begin{equation*}
\begin{aligned}
\mathbf{P}_3^{\prime\prime\prime}:\,\,& \underset{\bm{\tau}, \bm{G}, \bm{\phi}}{\text{minimize}}
& & \phi_3 + \text{Tr} (\bm{G}) + \phi_2 \\
& \text{subject to}
& & (\ref{eqn:tau}), (\ref{eqn:r12}), (\ref{eqn:gsd}), (\ref{eqn:vmatap}), (\ref{eqn:vrenergy}), (\ref{eqn:rs2ap}), (\ref{eqn:rs1ap2}).
\end{aligned}
\end{equation*}
\textcolor[rgb]{0.00,0.00,0.00}{\begin{prop}\label{pop3}
$\mathbf{P}_2^{\prime\prime\prime}$ is a convex problem.
\end{prop}
\begin{IEEEproof}
 The proof is similar to that of Proposition \ref{pop3}, which is omitted here.
\end{IEEEproof}}
As a result, the optimal solution $[\bm{\tau}^*,\bm{G}^*, \bm{\phi}^*]$ of Problem $\mathbf{P}_3^{\prime\prime\prime}$ can be obtained by using some known solution methods, such as the interior point method, etc.

\subsection{Global Optimum Analysis for Our Proposed Solution Method}
As is known, with SDR method, only when $\text{rank}(\bm{G}^*) = 1$, $[\bm{\tau}^*,\bm{G}^*, \bm{\phi}^*]$ is also the optimal solution of Problem $\mathbf{P}_3^{\prime\prime}$. In this case, the optimal  $[\bm{\tau}^*,\bm{\omega}^*,\mathbf{P}^*]$ can be derived accordingly. Therefore, the key question lies in the rank of $\bm{G}^*$. Fortunately, we also found that $\text{rank}(\bm{G}^*) = 1$ always holds for Problem $\mathbf{P}_3^{\prime\prime\prime}$, which means the global optimum of the primary Problem $\mathbf{P}_3$ also can be guaranteed by our adopted variable substitutions and SDR.

Now we analyse the rank of $\bm{G}^*$ for the minimum average power design with Theorem \ref{theo:vrank1mp}.

\begin{theorem}\label{theo:vrank1mp}
There exists an optimal $\bm{G}^*$ of Problem $\mathbf{P}_3^{\prime\prime\prime}$ such that $\text{rank}(\bm{G}^*) = 1$.
\end{theorem}
\begin{IEEEproof}
The proof can be found in Appendix \ref{App:T3}.
\end{IEEEproof}

\begin{corollary}\label{rmk:p3}
  The optimal solution to Problem $\mathbf{P}_3$ is guaranteed by using our proposed method.
\end{corollary}
\begin{IEEEproof}
The proof of Corollary 3 is similar to that of Corollary 1. $\mathbf{P}_3$, $\mathbf{P}_3^{\prime}$ and $\mathbf{P}_3^{\prime\prime}$ are equivalent to each other. Theorem \ref{theo:vrank1mp} declares that
$\mathbf{P}_3^{\prime\prime\prime}$ has a rank-one optimal solution. Therefore, the optimal solution to Problem $\mathbf{P}_3$ can always be found by using our proposed solution method.
\end{IEEEproof}

\section{Numerical Result \& Discussion}
In this section, we provide some numerical results to discuss the system performance of the optimized cooperative WPCN. For comparison, \textcolor[rgb]{0.00,0.00,0.00}{three} benchmark systems are also simulated. In the first benchmark system, \textcolor[rgb]{0.00,0.00,0.00}{i.e., random beamforming with optimized time assignment (RBOT)},  only time assignment is optimized and the power of $\rm{S}_1$ is \textcolor[rgb]{0.00,0.00,0.00}{randomly} allocated to its antennas. In the second benchmark system, \textcolor[rgb]{0.00,0.00,0.00}{i.e., optimized beamforming with random time assignment (OBRT),} only beamforming is optimized and random time assignment is adopted. \textcolor[rgb]{0.00,0.00,0.00}{In the third benchmark system,  i.e., random beamforming with random time assignment (RBRT), both beamforming and time assignment are randomly generated.}

In the simulations, we set $P_{\rm{S}_1}=2$Watt, $P_{\rm{S}_2}=0.2$Watt and $N_0=10^{-6}$Watt. Moreover, the minimal required rates of the two groups are set as $r_{\rm{S}_1}=0.5$bit/s and $r_{\rm{S}_2}=0.2$bit/s, respectively. The distances between the nodes are $d_{\rm{S}_1\rm{D}_1}  = 9$m, $d_{\rm{S}_1\rm{R}}  = 2$m, $d_{\rm{S}_2\rm{R}} = 10$m and $d_{\rm{R}\rm{D}_2}  = 20$m. A very weak direct link between $\rm{S}_2$ and $\rm{D}_2$ is assumed, which is with an equivalent distance as $d_{\rm{S}_2\rm{D}_2}  = 100$m. The pass loss exponential factor is 4. The number of antenna $N=4$ and the energy conversion efficiency $\eta=0.9$.
 These configurations will not change unless otherwise specified.

\subsection{Maximum WSR Performance}

\begin{figure}
\centering
\includegraphics[width=0.455\textwidth]{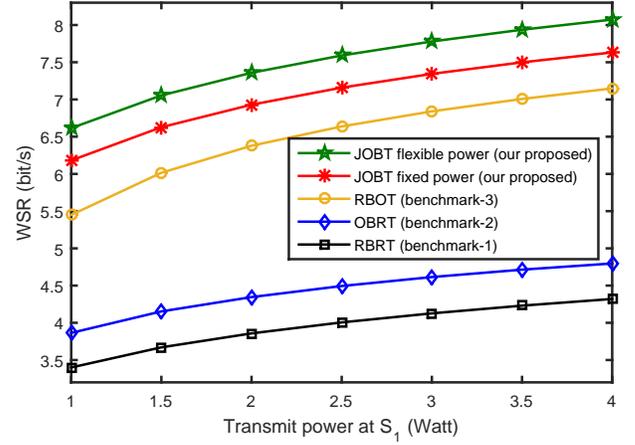}
\caption{\textcolor[rgb]{0.00,0.00,0.00}{System maximum WSR v.s. transmit power at $\rm{S}_1$}.}
\label{fig:WSRvsPS1}
\end{figure}
\begin{figure}
\centering
\includegraphics[width=0.455\textwidth]{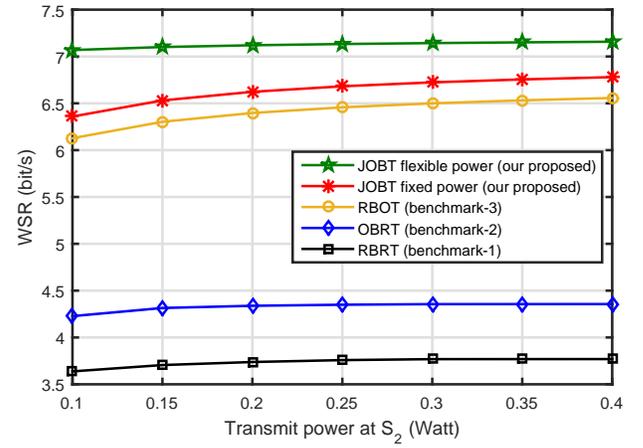}
\caption{\textcolor[rgb]{0.00,0.00,0.00}{System maximum WSR v.s. transmit power at $\rm{S}_2$}.}
\label{fig:WSRvsPS2}
\end{figure}

In Figure \ref{fig:WSRvsPS1} and Figure \ref{fig:WSRvsPS2}, the system WSR versus $P_{\rm{S}_1}$ and $P_{\rm{S}_2}$ are respectively plotted, where $\alpha_1=\alpha_2=1$. It can be seen that with the increment of $P_{\rm{S}_1}$ and $P_{\rm{S}_2}$, the WSRs of all \textcolor[rgb]{0.00,0.00,0.00}{five} systems increase. The reason is  a little bit straightforward, because more power will bring higher information rate. It also can be observed that
\textcolor[rgb]{0.00,0.00,0.00}{RBOT outperforms ORBT and RBRT, and RBRT achieves the lowest WSR among all systems}. This indicates that in the considered WPCN system, the time assignment has greater impact on the system performance than the beamforming at $\rm{S}_1$. The reason may be explained as follows. The beamforming design affects the system performance by energy transfer, which directly works on $\rm{R}$ and $\rm{D}_1$. Since the power transfer over wireless channels is faded seriously, its effects is relatively limited; while the time assignment works on all source and relay nodes, which adjusts the system resources more systematically. Therefore, time assignment has much greater impact on system performance \textcolor[rgb]{0.00,0.00,0.00}{and it is more} important in enhancing system performance. Besides, it is shown that compared with the fixed power constraints, flexible power configuration may greatly increase the system WSR. The performance gain between the system with  flexible power constraint and the one  with fixed power constraint is yielded by power allocation, which indicates that with power allocation at the two sources, the system WSR can be greatly improved.

\begin{figure}
\centering
\includegraphics[width=0.455\textwidth]{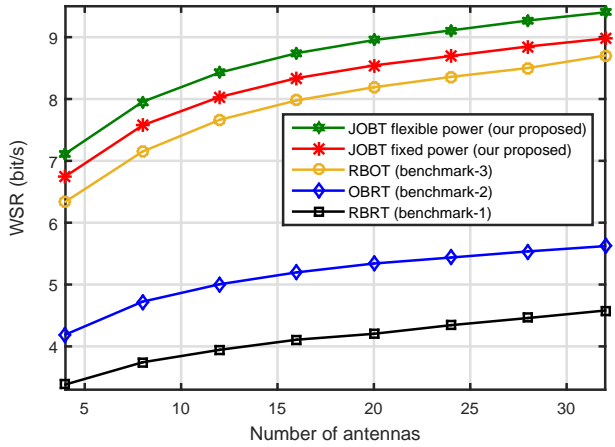}
\caption{\textcolor[rgb]{0.00,0.00,0.00}{System maximum WSR v.s. the number of antennas at $\rm{S}_1$}.}
\label{fig:WSRvsN}
\end{figure}
In Figure \ref{fig:WSRvsN}, the WSR is plotted versus the number of antennas of $\rm{S}_1$. One can see that as antenna number increases, the system WSR is also increasing. Moreover, it also shows that with the increment of the number of antennas, the increasing rate of the WSR roughly decreases, which means that increasing the number of antennas is able to enhance system WSR, but it cannot increase the system WSR infinitely.

\begin{figure}
\centering
\includegraphics[width=0.465\textwidth]{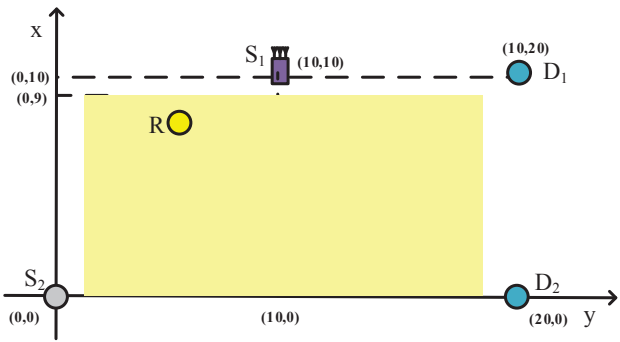}
\caption{\textcolor[rgb]{0.00,0.00,0.00}{Illustration of simulation topology on discussing the effect of relay position.}}
\label{fig:Sim}
\end{figure}
\begin{figure}
\centering
\includegraphics[width=0.455\textwidth]{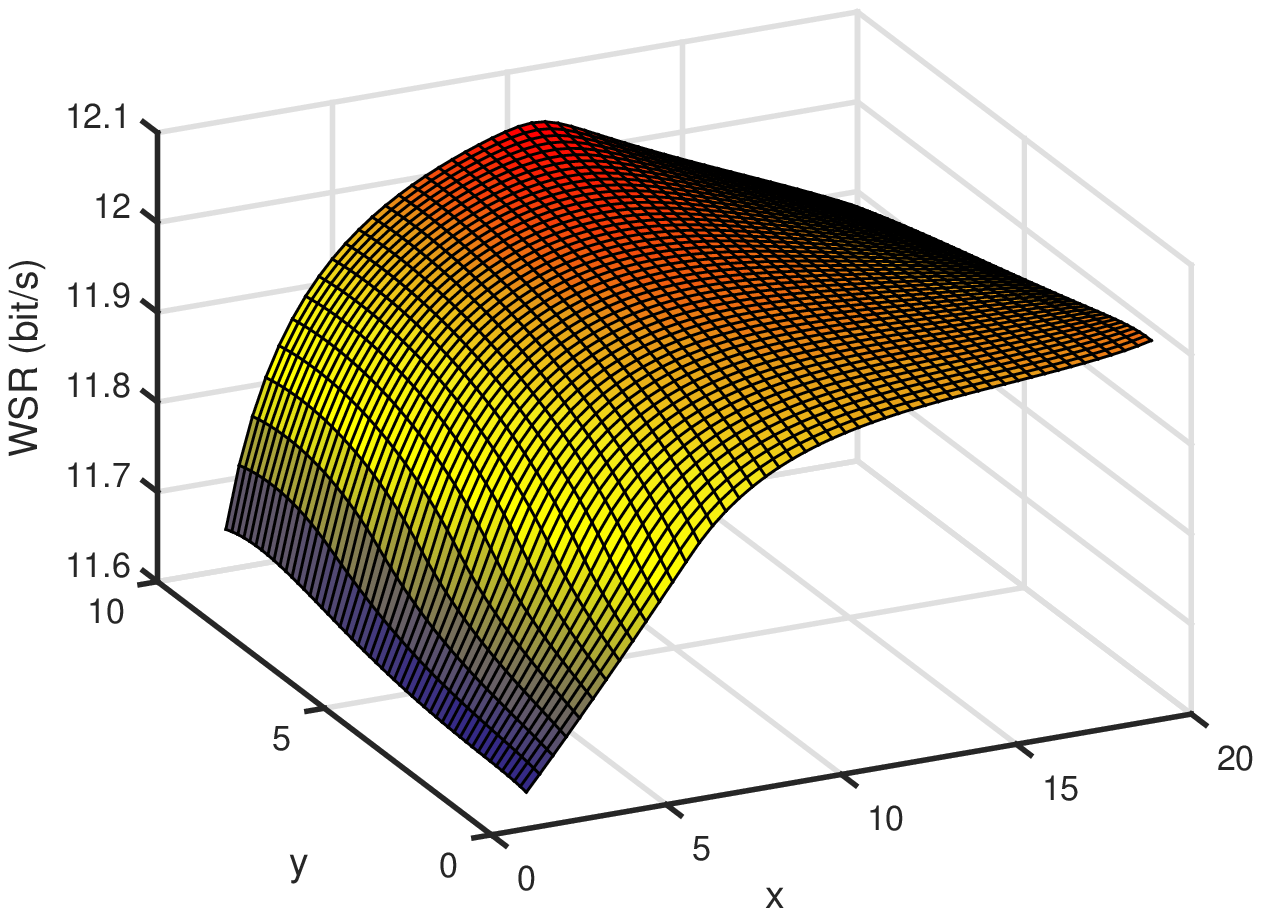}
\caption{System maximum WSR v.s. relay position for fixed power scenario.}
\label{fig:WSRvs3D}
\end{figure}

\begin{figure}
\centering
\includegraphics[width=0.455\textwidth]{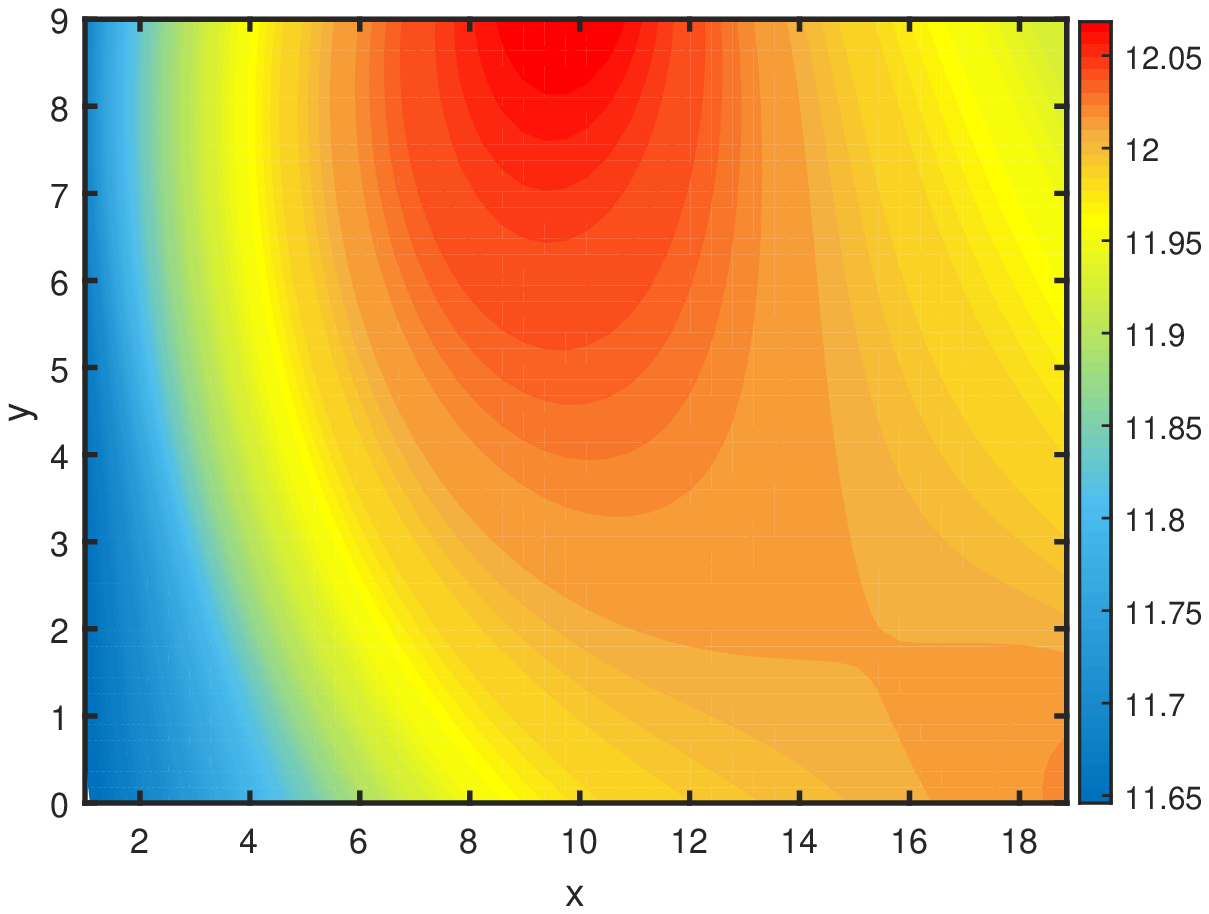}
\caption{Contour of the system maximum WSR v.s. relay position for fixed power scenario.}
\label{fig:WSRcontours}
\end{figure}
%
%\begin{figure*}
%\begin{minipage}[t]{0.5\linewidth}
%\centering
%\includegraphics[width=0.95\textwidth]{WSRvs3D.eps}
%\caption{System maximum WSR v.s. relay position for fixed\protect\\ power scenario.}
%\label{fig:WSRvs3D}
%\end{minipage}%
%\begin{minipage}[t]{0.5\linewidth}
%\centering
%\includegraphics[width=0.95\textwidth]{WSRvsCountour.eps}
%\caption{Contour of the system maximum WSR v.s. relay position for fixed power scenario.}
%\label{fig:WSRcontours}
%\end{minipage}
%\end{figure*}

%\begin{figure*}
%\begin{minipage}[t]{0.5\linewidth}
%\centering
%\includegraphics[width=0.95\textwidth]{WSRvs3D2.eps}
%\caption{\textcolor[rgb]{0.00,0.00,0.00}{System maximum WSR v.s. relay position for flex-\protect\\ible power scenario}.}
%\label{fig:WSRvs3D2}
%\end{minipage}%
%\begin{minipage}[t]{0.5\linewidth}
%\centering
%\includegraphics[width=0.95\textwidth]{WSRvsCountour2.eps}
%\caption{\textcolor[rgb]{0.00,0.07,1.00}{Contour of the system maximum WSR v.s. relay position for flexible power scenario}.}
%\label{fig:WSRcontours2}
%\end{minipage}\vspace{-0.1in}
%\end{figure*}

\begin{figure}
\centering
\includegraphics[width=0.455\textwidth]{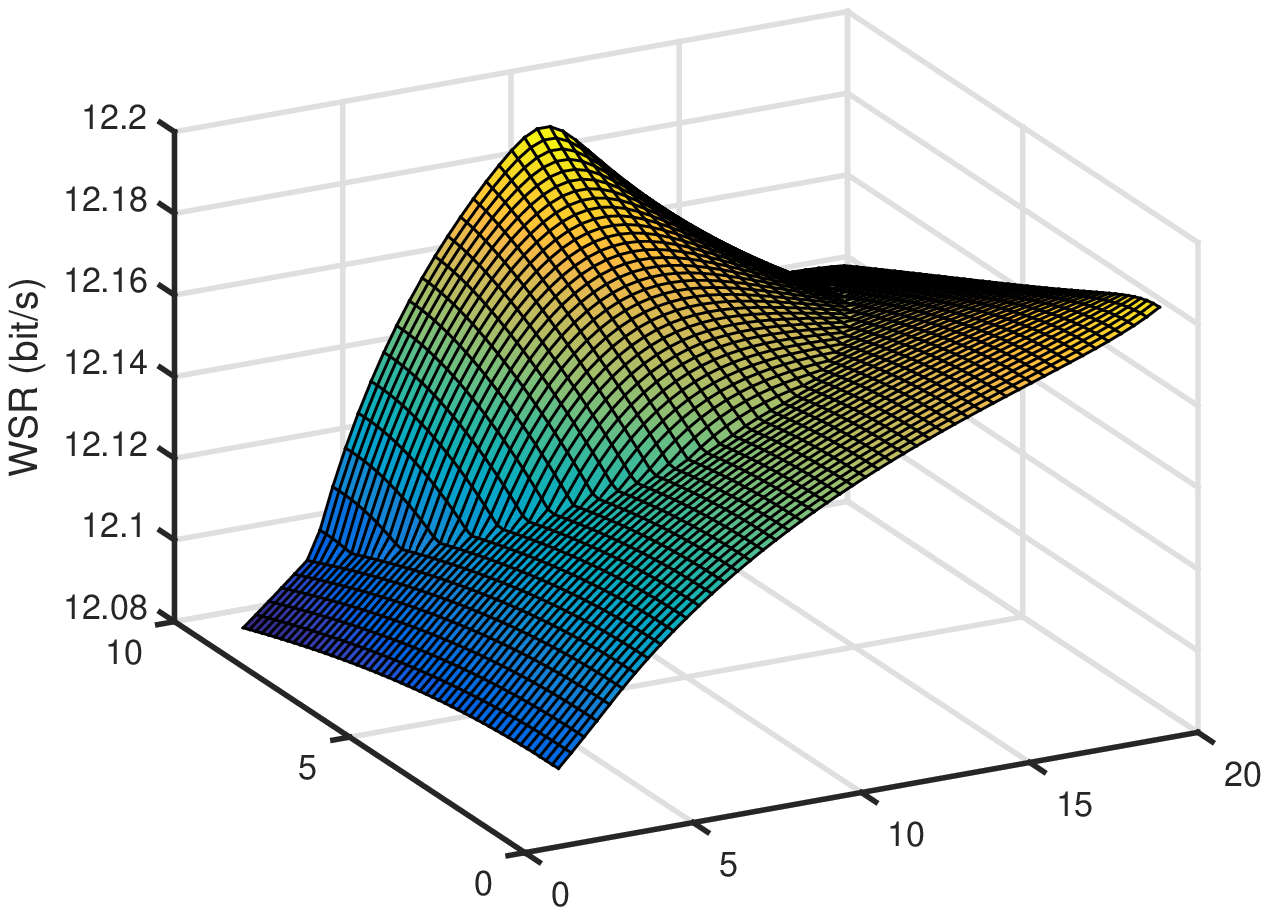}
\caption{\textcolor[rgb]{0.00,0.00,0.00}{System maximum WSR v.s. relay position for flexible power scenario}.}
\label{fig:WSRvs3D2}
\end{figure}

\begin{figure}
\centering
\includegraphics[width=0.455\textwidth]{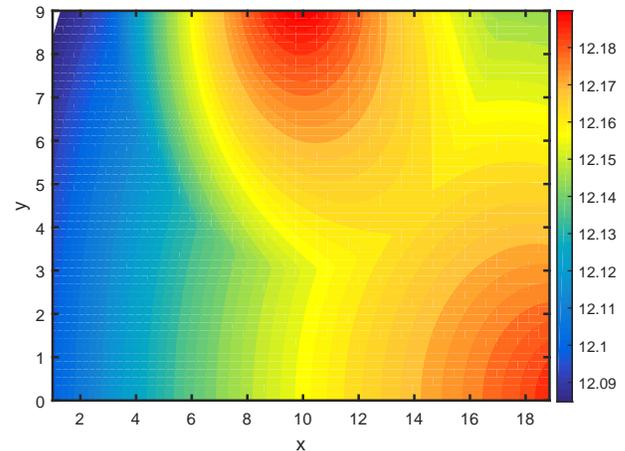}
\caption{\textcolor[rgb]{0.00,0.07,1.00}{Contour of the system maximum WSR v.s. relay position for flexible power scenario}.}
\label{fig:WSRcontours2}
\end{figure}
To discuss the effect of relay position on system performance, we also simulate the WSR versus different relay locations. In the simulations, \textcolor[rgb]{0.00,0.00,0.00}{we consider a network topology as shown in Figure \ref{fig:Sim}}, where $\rm{S}_2$ is located at the origin of the coordination on the $x-y$ plane, ${\rm D}_2$ is located at the point with coordinate $(x = 10, y = 0)$, $\rm{S}_1$ is positioned at $(x = 10, y = 10)$ and ${\rm D}_2$ is placed at $(x = 20, y = 10)$. The position of $\rm R$ is changed within the region of $1 \leq x \leq 19$ and $0 \leq y \leq 9$. From the result in \textcolor[rgb]{0.00,0.00,0.00}{Figure \ref{fig:WSRvs3D} and Figure \ref{fig:WSRvs3D2}}, it can be seen that the relay should be positioned closer to $\rm{S}_1$ for higher system WSR. When it is closer to $\rm{S}_2$, the system achieves relatively low WSR. In order to show this more clearly, the contour lines \textcolor[rgb]{0.00,0.00,0.00}{associated withFigure \ref{fig:WSRvs3D} and Figure \ref{fig:WSRvs3D2}} are plotted in \textcolor[rgb]{0.00,0.00,0.00}{Figure \ref{fig:WSRcontours} and Figure \ref{fig:WSRcontours2}, respectively,} which also shows that when the relay is placed closer to $\rm{S}_1$ or $\rm{D}_2$, a relatively high WSR can be achieved. This result can be applied to relay deployment or relay section in the practical cooperative WPCNs.

\subsection{Minimal Power performance simulations}
\begin{figure}
\centering
\includegraphics[width=0.455\textwidth]{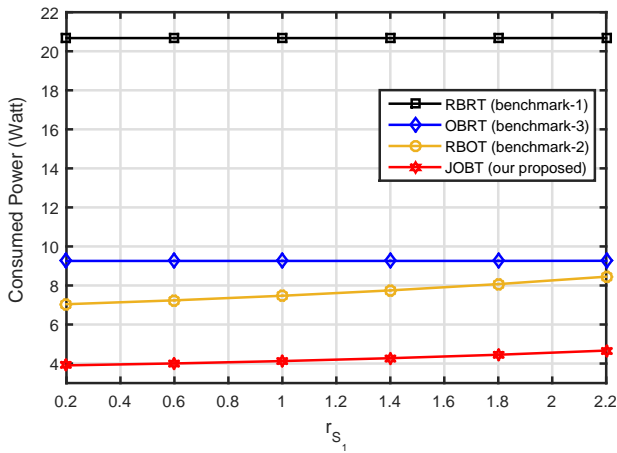}
\caption{\textcolor[rgb]{0.00,0.00,0.00}{Minimum consumed power v.s. the rate threshold $r_{\rm{S}_1}$}.}
\label{fig:Powervsr1}
\end{figure}

\begin{figure}
\centering
\includegraphics[width=0.455\textwidth]{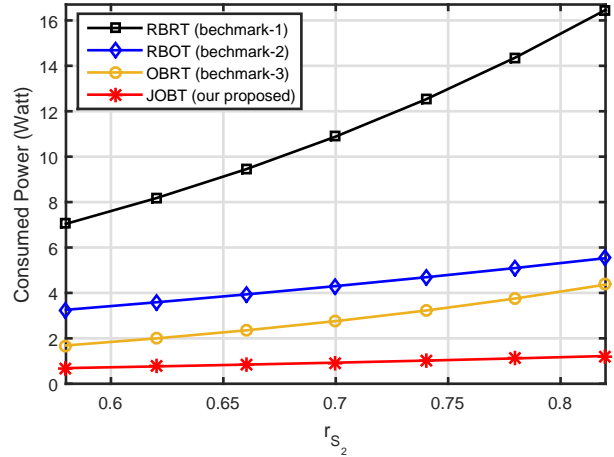}
\caption{\textcolor[rgb]{0.00,0.00,0.00}{Minimum consumed power v.s. the rate threshold $r_{\rm{S}_2}$}.}
\label{fig:Powervsr2}
\end{figure}

In Figure \ref{fig:Powervsr1} and Figure \ref{fig:Powervsr2}, the system minimum consumed power of our proposed method and the benchmark systems, \textcolor[rgb]{0.00,0.00,0.00}{i.e., RBOT, OBRT and RBRT} are plotted versus $r_{\rm{S}_1}$ and $r_{\rm{S}_2}$, respectively. It can be seen that, with the increment of $r_{\rm{S}_1}$ and $r_{\rm{S}_2}$, the total consumed power of \textcolor[rgb]{0.00,0.00,0.00}{four} systems increase, since to meet the higher data rate requirements of the two groups, more power are required. It also shows that the minimum consumed power of \textcolor[rgb]{0.00,0.00,0.00}{the four} systems increase more quickly with the increment of $r_{\rm{S}_2}$ than that with the increment of $r_{\rm{S}_1}$. This indicates that to meet data rate requirement of group 2 consumes more power. The reason is that the available power at $\rm R$ is transferred from ${\rm S}_1$ and \textcolor[rgb]{0.00,0.00,0.00}{during the energy transfer} some energy is lost  due to path loss fading.

\begin{figure}
\centering
\includegraphics[width=0.455\textwidth]{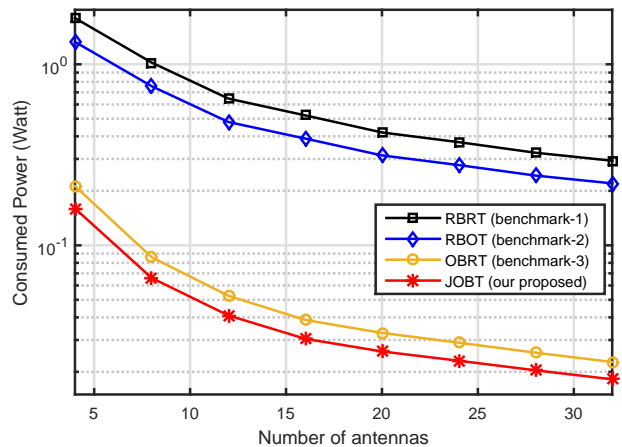}
\caption{\textcolor[rgb]{0.00,0.00,0.00}{Minimum consumed power v.s. the number of antennas at ${\rm{S}_1}$}.}
\label{fig:PowvsN}
\end{figure}

In Figure \ref{fig:PowvsN}, the system minimum consumed power is plotted versus the number of antennas of $\rm{S}_1$. It can be seen that as antenna number increases, the total consumed power is reduced. However, with the increment of the number of antennas, the decreasing rate of the total consumed power decreases, which means that increasing the number of antennas is capable of decrease the system total consumed power, but it cannot decrease the system total consumed power infinitely.

%\begin{figure*}
%\begin{minipage}[t]{0.5\linewidth}
%\centering
%\includegraphics[width=0.95\textwidth]{Powervs3D.eps}
%\caption{Minimum consumed power v.s. the relay position.}\label{fig:Powvs3D}
%\end{minipage}%
%\begin{minipage}[t]{0.5\linewidth}
%\centering
%\includegraphics[width=0.95\textwidth]{PowervsCountour.eps}
%\caption{Contour of the Minimum consumed power v.s. the relay position.}\label{fig:Powvscontours}
%  %where the transmit power constraint of each cell is $46 \rm dBm$ and the required SINR of each low-mobility user is $10 {\rm dB}$.\protect
%\end{minipage}\vspace{-0.1in}
%\end{figure*}
\begin{figure}
\centering
\includegraphics[width=0.455\textwidth]{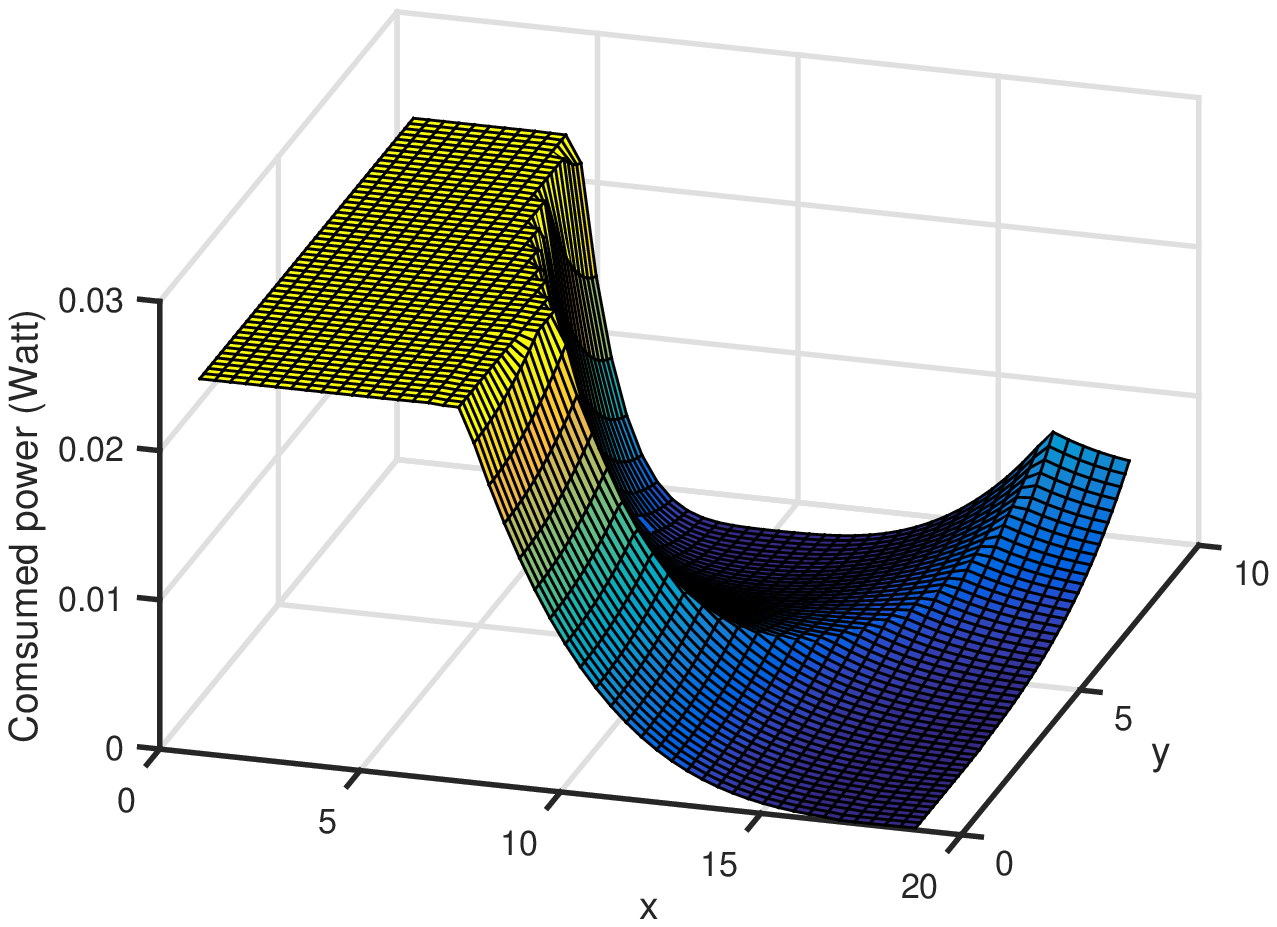}
\caption{Minimum consumed power v.s. the relay position.}
\label{fig:Powvs3D}
\end{figure}

\begin{figure}
\centering
\includegraphics[width=0.455\textwidth]{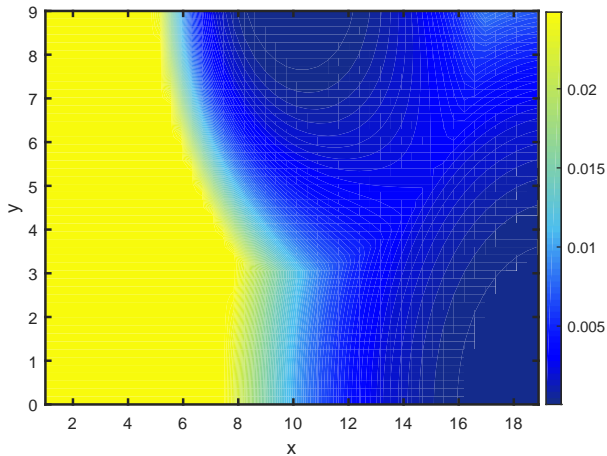}
\caption{Contour of the Minimum consumed power v.s. the relay position.}
\label{fig:Powvscontours}
\end{figure}
To discuss the effect of relay position on the system total consumed power, in Figure \ref{fig:Powvs3D}, we simulate the the minimum consumed power versus different relay locations. In the simulations, we \textcolor[rgb]{0.00,0.00,0.00}{also consider the topology as shown in Figure \ref{fig:Sim}}. The position of $\rm R$ is changed within the region of $1 \leq x \leq 19$ and $0 \leq y \leq 9$. From Figure \ref{fig:Powvs3D}, it can be seen that the relay should be positioned closer to $\rm{S}_1$ or $\rm{D}_2$ for achieving a lower total consumed power. When it is closer to $\rm{S}_2$, the system achieves relatively high lower total consumed power. In order to show this more clearly, the contour lines are plotted in Figure \ref{fig:Powvscontours}. \textcolor[rgb]{0.00,0.00,0.00}{The results also can be used as a reference for relay deployment or relay section in the practical cooperative WPCNs.}

\section{Conclusion}
This paper studied the optimal resource allocation for the WPCN with group cooperation. We introduced \textcolor[rgb]{0.00,0.00,0.00}{energy cooperation and time sharing} between the two groups, so that both groups could fulfill their expected information delivering. To explore the system performance limits, we formulated  optimization problems to maximize the system WSR and minimize its total consumed power by jointly optimizing the time assignment, power allocation, and SWIPT beamforming vectors under the available power constraint and the QoS requirement constraints of  both groups. We solved the problems by using proper variable substitutions and the SDR method.  We theoretically proved that our proposed solution methods can guarantee the global optimal solutions.  Numerical results were provided to discuss the system performance behaviors. \textcolor[rgb]{0.00,0.00,0.00}{It showed that in such a group cooperation-aware WPCN, optimal time assignment has the most great effect on the system performance than other factors. Besides, the effects of relay position on system performances are also discussed via simulations}.

In future systems, some advanced technologies, such as network coding\cite{Fan1}, OFDM\cite{Fan2} and cognitive sensing, etc may be instigated into WPCNs to enhance the system performance. Besides,  such kind of WPCNs also may be extended to high-speed railway scenarios\cite{Fan3} for more widely application.

\begin{appendices}
\section{{The Proof of Theorem 1}}\label{App:T1}
First, we consider the following Problem $\mathbf{Q}_1$,
\begin{flalign}
\mathbf{Q}_1:\,\,\,\underset{\mathbf{U}}{\text{minimize}}\,\,& \text{Tr}(\mathbf{U})\\
\text{subject to}\,\,\,& \phi_4^* \leq \eta P_{\rm{S}_1} \text{Tr} (P_{\rm{S}_1}\mathbf{U} \mathbf{h}_{\rm{S}_1 \rm{R}} \mathbf{h}_{\rm{S}_1 \rm{R}}^H ), \,\,\mathbf{U} \succeq 0,\nonumber\\
&R_{\rm{S}_1}^* =\tau_1^* \mathcal{C}\left(\frac{\text{Tr} (P_{\rm{S}_1}\mathbf{U} \mathbf{h}_{\rm{S}_1 \rm{D}_1} \mathbf{h}_{\rm{S}_1 \rm{D}_1}^H )}{N_0 \tau_1^*}\right)\nonumber\\
&\quad\quad+\textcolor[rgb]{0.00,0.00,0.00}{\tau_2}^*\mathcal{C}\left( \frac{P_{\rm{S}_1} \|\mathbf{h}_{\rm{S}_1 \rm{D}_1}\|^2}{N_0}\right), \nonumber
\end{flalign}
where $\tau_1^*$, $\textcolor[rgb]{0.00,0.00,0.00}{\tau_2}^*$, $\phi_4^*$and $R_{\rm{S}_1}^*$ are  optimal solutions of Problem $\mathbf{P}_1^{\prime\prime\prime}$.
Further, it can be equivalently transformed into
\begin{flalign}
\mathbf{Q}_1^{\prime}:\,\,\underset{\mathbf{U}}{\text{minimize}}\,\,&\text{Tr}(\mathbf{U})\\
\text{subject to}\,\,&P_{\rm{S}_1}\text{Tr} (\mathbf{U} \mathbf{h}_{\rm{S}_1 \rm{R}} \mathbf{h}_{\rm{S}_1 \rm{R}}^H ) \geq \frac{\phi_4^*}{\eta},\nonumber \\
& P_{\rm{S}_1}\text{Tr} (\mathbf{U} \mathbf{h}_{\rm{S}_1 \rm{D}_1} \mathbf{h}_{\rm{S}_1 \rm{D}_1}^H ) = N_0 \tau_1^* \beta,\,\,\mathbf{U} \succeq 0,\nonumber
\end{flalign}
where $\beta=\left( 2^{\frac{R_{\rm{S}_1}^* - \textcolor[rgb]{0.00,0.00,0.00}{\tau_2}^*\mathcal{C}\big( \frac{P_{\rm{S}_1}  \|\mathbf{h}_{\rm{S}_1 \rm{D}_1}\|^2}{N_0}\big)}{\tau_1^*}} - 1 \right)$.
According to Lemma \ref{lem:vrank1}, Problem $\mathbf{Q}_1$ has an optimal solution $\mathbf{U}^*$ which satisfies that
\begin{equation*}
  \text{rank}^2(\mathbf{U}^*) \leq 2.
\end{equation*}
Moreover, since $\text{rank}(\mathbf{U}^*) \neq 0$, $\text{rank}(\mathbf{U}^*) = 1$.

%First, we need to establish that Problem $\mathbf{P}_1^{\prime\prime\prime}$ and Problem $\mathbf{Q}_1$ have the same optimal solution.
Let $[\bm{\tau}^*,\bm{\digamma}^*, \phi_4^*]$ be the optimal solution of Problem $\mathbf{P}_1^{\prime\prime\prime}$. It can be inferred that $\bm{\digamma}^*$ is a feasible solution of Problem $\mathbf{Q}_1$. The reason is that $[\bm{\tau}^*,\bm{\digamma}^*, \phi_4^*]$ also satisfy the constraints (\ref{eqn:vrs1}) and (\ref{eqn:vrenergy}). The optimal value of Problem $\mathbf{Q}_1$ associated with $\mathbf{U}^*$ must be smaller than that associated with any other feasible solution. Therefore, $\text{Tr}(\mathbf{U}^*) \leq \text{Tr}(\bm{\digamma}^*) \leq \tau_1^* P_{\rm{S}_1}$.

If we construct a new tuple $[\bm{\tau}^*,\mathbf{U}^*, \phi_4^*]$, then it satisfy all constraints of Problem $\mathbf{P}_1^{\prime\prime\prime}$, which means it is a feasible solution of Problem $\mathbf{P}_1^{\prime\prime\prime}$. Since the objective function of Problem $\mathbf{P}_1^{\prime\prime\prime}$ is only related to $\bm{\tau}$ and $\phi_4$, $[\bm{\tau}^*,\mathbf{U}^*, \phi_4^*]$ and $[\bm{\tau}^*,\bm{\digamma}^*, \phi_4^*]$ yield the same value of Problem $\mathbf{P}_1^{\prime\prime\prime}$, which means that $[\bm{\tau}^*,\mathbf{U}^*, \phi_4^*]$ is also an optimal solution of Problem $\mathbf{P}_1^{\prime\prime\prime}$.
Since we have proved that $\text{rank}(\mathbf{U}^*) = 1$, it can be concluded that $\mathbf{P}_1^{\prime\prime\prime}$ has an optimal rank-one solution.
\section{{The Proof of Theorem 2}}\label{App:T2}
First, we consider the following Problem $\mathbf{Q}_2$,
\begin{flalign}
\mathbf{Q}_2:\,\,\underset{\mathbf{U}}{\text{minimize}}\,\,
&\text{Tr}(\mathbf{U})\\
\text{subject to}\,\,&\phi_4^* \leq \eta P_{\rm{S}_1}^{(1)}\text{Tr} (\mathbf{U} \mathbf{h}_{\rm{S}_1 \rm{R}} \mathbf{h}_{\rm{S}_1 \rm{R}}^H ), \\
&R_{\rm{S}_1}^* = \tau_1^* \mathcal{C} \left( \frac{P_{\rm{S}_1}^{(1)}\text{Tr} (\mathbf{U} \mathbf{h}_{\rm{S}_1 \rm{D}_1} \mathbf{h}_{\rm{S}_1 \rm{D}_1}^H )}{N_0 \tau_1^*}\right)\nonumber\\
&+ \textcolor[rgb]{0.00,0.00,0.00}{\tau_2}^*\mathcal{C}\left( \frac{\phi_2^* \|\mathbf{h}_{\rm{S}_1 \rm{D}_1}\|^2}{N_0 \textcolor[rgb]{0.00,0.00,0.00}{\tau_2}^*}\right), \\
& \mathbf{U} \succeq 0,
\end{flalign}
where $\tau_1^*$, $\textcolor[rgb]{0.00,0.00,0.00}{\tau_2}^*$, $\phi_4^*$, $\phi_2^*$ and $R_{\rm{S}_1}^*$ are  optimal solutions of Problem $\mathbf{P}_2^{\prime\prime\prime}$.
Problem $\mathbf{Q}_2$ is equivalently transformed into Problem $\mathbf{Q}_2^{\prime}$,
\begin{flalign}
\mathbf{Q}_2^{\prime}:\,\,\underset{\mathbf{U}}{\text{minimize}}
\,\,&\text{Tr}(\mathbf{U})\\
\text{subject to}\,\,&P_{\rm{S}_1}^{(1)}\text{Tr} (\mathbf{U} \mathbf{h}_{\rm{S}_1 \rm{R}} \mathbf{h}_{\rm{S}_1 \rm{R}}^H ) \geq \frac{\phi_4^*}{\eta}, \\
&P_{\rm{S}_1}^{(1)}\text{Tr} (\mathbf{U} \mathbf{h}_{\rm{S}_1 \rm{D}_1} \mathbf{h}_{\rm{S}_1 \rm{D}_1}^H ) =\xi, \,\,\mathbf{U} \succeq 0,
\end{flalign}
where $\xi=N_0 \tau_1^* \Bigg( 2^{\frac{R_{\rm{S}_1}^* - \textcolor[rgb]{0.00,0.00,0.00}{\tau_2}^*\mathcal{C}\left(\frac{\phi_2^* \|\mathbf{h}_{\rm{S}_1 \rm{D}_1}\|^2}{N_0 \textcolor[rgb]{0.00,0.00,0.00}{\tau_2}^*}\right)}{\tau_1^*}} - 1 \Bigg)$.
According to Lemma \ref{lem:vrank1}, Problem $\mathbf{Q}_2$ has an optimal solution $\mathbf{U}^*$ which satisfies that
\begin{equation*}
  \text{rank}^2(\mathbf{U}^*) \leq 2.
\end{equation*}
Since $\text{rank}(\mathbf{U}^*) \neq 0$, we conclude that $\text{rank}(\mathbf{U}^*) = 1$.

%First, we need to establish that Problem $\mathbf{P}_1^{\prime\prime\prime}$ and Problem $\mathbf{Q}_1$ have the same optimal solution.
Let $[\bm{\tau}^*,\bm{\textcolor[rgb]{0.00,0.00,0.00}{G}}^*, \bm{\phi}^*]$ be the optimal solution of Problem $\mathbf{P}_2^{\prime\prime\prime}$. It can be inferred that $\bm{\textcolor[rgb]{0.00,0.00,0.00}{G}}^*$ is a feasible solution of Problem $\mathbf{Q}_2$. The reason is that $[\bm{\tau}^*,\bm{\textcolor[rgb]{0.00,0.00,0.00}{G}}^*, \bm{\phi}^*]$ also satisfy the constraints (\ref{eqn:vrenergy}) and (\ref{eqn:rs1ap2}). The optimal value of Problem $\mathbf{Q}_2$ associated with $\mathbf{U}^*$ must be smaller than that associated with any other feasible solution. Therefore, $\text{Tr}(\mathbf{U}^*) \leq \text{Tr}(\bm{\textcolor[rgb]{0.00,0.00,0.00}{G}}^*) \leq \frac{\phi_1^*}{P_{\rm{S}_1}^{(1)}}$.

If we construct a new tuple $[\bm{\tau}^*,\mathbf{U}^*, \bm{\phi}^*]$, then it satisfy all constraints of Problem $\mathbf{P}_2^{\prime\prime\prime}$, which means it is a feasible solution of Problem $\mathbf{P}_2^{\prime\prime\prime}$. Since the objective function of Problem $\mathbf{P}_2^{\prime\prime\prime}$ is only related to $\bm{\tau}$ and $\bm{\phi}$, $[\bm{\tau}^*,\mathbf{U}^*, \bm{\phi}^*]$ and $[\bm{\tau}^*,\bm{\textcolor[rgb]{0.00,0.00,0.00}{G}}^*, \bm{\phi}^*]$ yield the same value of Problem $\mathbf{P}_2^{\prime\prime\prime}$, which means that $[\bm{\tau}^*,\mathbf{U}^*, \bm{\phi}^*]$ is also an optimal solution of Problem $\mathbf{P}_2^{\prime\prime\prime}$.
Since we have proved that $\text{rank}(\mathbf{U}^*) = 1$, we conclude that $\mathbf{P}_2^{\prime\prime\prime}$ has an optimal rank-one solution.
\section{{The Proof of Theorem 3}}\label{App:T3}
First, we apply the substitution \begin{equation}\label{eqn:vsub}
  \text{Tr}(\bm{\textcolor[rgb]{0.00,0.00,0.00}{G}}) = t,
\end{equation}
on Problem $\mathbf{P}_3^{\prime\prime\prime}$ and get an equivalent Problem $\mathbf{\Delta}$,
\begin{equation*}
\begin{aligned}
\mathbf{\Delta}:\,\,& \underset{\bm{\tau}, \bm{\textcolor[rgb]{0.00,0.00,0.00}{G}}, \bm{\phi}, t}{\text{minimize}}
& & \phi_3 + tP_{\rm{S}_1}^{(1)} + \phi_2 \\
& \text{subject to}
& & (\ref{eqn:tau}), (\ref{eqn:r12}), (\ref{eqn:vsd}), (\ref{eqn:vrenergy}), (\ref{eqn:rs2ap}), (\ref{eqn:vmatap}), (\ref{eqn:rs1ap2}), (\ref{eqn:vsub}).
\end{aligned}
\end{equation*}

Next, we consider the following Problem $\mathbf{Q}_3$,
\begin{flalign}
\mathbf{Q}_3:\,\,\underset{\mathbf{U}}{\text{minimize}}\,\,&\text{Tr}(\mathbf{U})\\
\text{subject to}\,\,&\phi_4^* \leq \eta P_{\rm{S}_1}^{(1)}\text{Tr} (\mathbf{U} \mathbf{h}_{\rm{S}_1 \rm{R}} \mathbf{h}_{\rm{S}_1 \rm{R}}^H ), \\
&R_{\rm{S}_1}^* = \tau_1^* \mathcal{C}\left( \frac{P_{\rm{S}_1}^{(1)}\text{Tr} (\mathbf{U} \mathbf{h}_{\rm{S}_1 \rm{D}_1} \mathbf{h}_{\rm{S}_1 \rm{D}_1}^H )}{N_0 \tau_1^*}\right)\nonumber\\
&\quad+\textcolor[rgb]{0.00,0.00,0.00}{\tau_2}^*\mathcal{C}\left( \frac{\phi_2^* \|\mathbf{h}_{\rm{S}_1 \rm{D}_1}\|^2}{N_0 \textcolor[rgb]{0.00,0.00,0.00}{\tau_2}^*}\right), \\
&\text{Tr}(\mathbf{U}) = t^*,
\,\,\mathbf{U} \succeq 0,
\end{flalign}
where $\tau_1^*$, $\textcolor[rgb]{0.00,0.00,0.00}{\tau_2}^*$, $\phi_4^*$, $\phi_2^*$, $t^*$ and $R_{\rm{S}_1}^*$ are optimal solutions of Problem $\mathbf{\Delta}$.
Problem $\mathbf{Q}_3$ is equivalently transformed into Problem $\mathbf{Q}_3^{\prime}$,
\begin{flalign}
\mathbf{Q}_3^{\prime}:\,\,\underset{\mathbf{U}}{\text{minimize}}\,\,&\text{Tr}(\mathbf{U})\\
\text{subject to}\,\,
&\text{Tr} (\mathbf{U} \mathbf{h}_{\rm{S}_1 \rm{R}} \mathbf{h}_{\rm{S}_1 \rm{R}}^H ) \geq \frac{\phi_4^*}{\eta P_{\rm{S}_1}^{(1)}}, \\
& P_{\rm{S}_1}^{(1)}\text{Tr} (\mathbf{U} \mathbf{h}_{\rm{S}_1 \rm{D}_1} \mathbf{h}_{\rm{S}_1 \rm{D}_1}^H ) = \xi, \\
& \text{Tr}(\mathbf{U}) = t^*,\\
&\mathbf{U} \succeq 0.
\end{flalign}
%where $\xi=N_0 \tau_1^* \left( 2^{\frac{R_{\rm{S}_1}^* - \textcolor[rgb]{0.00,0.00,0.00}{\tau_2}^*\mathcal{C}\left( \frac{\phi_2^* \|\mathbf{h}_{\rm{S}_1 \rm{D}_1}\|^2}{N_0 \textcolor[rgb]{0.00,0.00,0.00}{\tau_2}^*}\right)}{\tau_1^*}} - 1 \right)$.
According to Lemma \ref{lem:vrank1}, Problem $\mathbf{Q}_3$ has an optimal solution $\mathbf{U}^*$ which satisfies that
\begin{equation*}
  \text{rank}^2(\mathbf{U}^*) \leq 3.
\end{equation*}
Since $\text{rank}(\mathbf{U}^*) \neq 0$, we conclude that $\text{rank}(\mathbf{U}^*) = 1$.

%First, we need to establish that Problem $\mathbf{P}_1^{\prime\prime\prime}$ and Problem $\mathbf{Q}_1$ have the same optimal solution.
Let $[\bm{\tau}^*,\bm{\textcolor[rgb]{0.00,0.00,0.00}{G}}^*, \bm{\phi}^*, t^*]$ be the optimal solution of Problem $\mathbf{G}$. It can be inferred that $\bm{\textcolor[rgb]{0.00,0.00,0.00}{G}}^*$ is a feasible solution of Problem $\mathbf{Q}_3$. The reason is that $[\bm{\tau}^*,\bm{\textcolor[rgb]{0.00,0.00,0.00}{G}}^*, \bm{\phi}^*, t^*]$ also satisfy the constraints (\ref{eqn:vrenergy}), (\ref{eqn:rs1ap2}) and (\ref{eqn:vsub}). The optimal value of Problem $\mathbf{Q}_3$ associated with $\mathbf{U}^*$ must be smaller than that associated with any other feasible solution. Therefore, $\text{Tr}(\mathbf{U}^*) \leq \text{Tr}(\bm{\textcolor[rgb]{0.00,0.00,0.00}{G}}^*) \leq \phi_1^*$.

If we construct a new tuple $[\bm{\tau}^*,\mathbf{U}^*, \bm{\phi}^*, t^*]$, then it satisfy all constraints of Problem $\mathbf{G}$, which means it is a feasible solution of Problem $\mathbf{G}$. Since the objective function of Problem $\mathbf{G}$ is only related to $\bm{\tau}$, $\bm{\phi}$ and $t$. $[\bm{\tau}^*,\mathbf{U}^*, \bm{\phi}^*, t^*]$ and $[\bm{\tau}^*,\bm{\textcolor[rgb]{0.00,0.00,0.00}{G}}^*, \bm{\phi}^*, t^*]$ yield the same value of Problem $\mathbf{G}$, which means that $[\bm{\tau}^*,\mathbf{U}^*, \bm{\phi}^*, t^*]$ is also an optimal solution of Problem $\mathbf{\Delta}$.
Since we have proved that $\text{rank}(\mathbf{U}^*) = 1$, we conclude that $\mathbf{\Delta}$ has an optimal rank-one solution. We also know that Problem $\mathbf{P}_3^{\prime\prime\prime}$ is equivalent to Problem $\mathbf{\Delta}$. So $\mathbf{P}_3^{\prime\prime\prime}$ also has an optimal rank-one solution.
\end{appendices}

%\section*{Acknowledgment}
%This work was supported by the Beijing Natural
%Science Foundation, no. 4162049, partly by the National Basic Research
%Program of China (973 Program), no. 2012CB316100(2) and also by the Open Research Fund of National Mobile
%Communications Research Laboratory, Southeast University (no. 2014D03).


% http://www.ctan.org/tex-archive/macros/latex/contrib/supported/IEEEtran/bibtex
%\bibliographystyle{IEEEtran.bst}

\end{document}